\documentclass[10pt,journal,compsoc]{IEEEtran}
\ifCLASSOPTIONcompsoc
  \usepackage[nocompress]{cite}
\else
  \usepackage{cite}
\fi
\ifCLASSINFOpdf
\else
\fi

\usepackage{bm}
\usepackage{amsmath}
\usepackage{subfigure}
\usepackage{array,graphicx}
\usepackage{subfloat}
\usepackage{graphicx}
\usepackage{cite}
\usepackage{algorithm}
\usepackage{multirow}
\usepackage[square, comma, sort&compress, numbers]{natbib}
\usepackage{mathrsfs}
\usepackage{caption2}
\usepackage{color}
\usepackage{booktabs}
\usepackage{times}
\usepackage{epsfig}
\usepackage{graphicx}
\usepackage{amsmath}
\usepackage{amssymb}
\usepackage{bbm}
\usepackage{multirow}
\usepackage{eqparbox,array}
\usepackage{makecell}

\begin{document}
%
\title{Entity-Graph Enhanced Cross-Modal Pretraining for Instance-level Product Retrieval}
%
%
%
%

\author{Xiao~Dong*, \quad 
        Xunlin~Zhan*, \quad
        Yunchao~Wei,\quad
        Xiaoyong~Wei,\quad
        Yaowei~Wang,\quad
        Minlong~Lu,\quad
        Xiaochun~Cao,\quad
        and\quad~Xiaodan~Liang,~\IEEEmembership{Member,~IEEE}
 \IEEEcompsocitemizethanks{\IEEEcompsocthanksitem X. Dong is with the School
of Artificial Intelligence, Zhuhai Campus, Sun Yat-Sen University, Zhuhai, P.R. China, 519082. X.L. Zhan and X.D. Liang  are with the School of Intelligent Systems Engineering, Shenzhen Campus, Sun Yat-Sen University, Shenzhen,  P.R. China, 528406. Y.C. Wei is with the School of Computer and Information Technology,  Beijing Jiaotong University,  Beijing,  P.R. China, 100044. X.Y. Wei and Y.W. Wang  are with PengCheng Laboratory, Shenzhen,  P.R. China, 322099. M.L. Lu is with Alibaba Group, Hangzhou,  P.R. China, 322099.  X.C. Cao is with School of Cyber Science and Technology, Shenzhen Campus, Sun Yat-sen University, Shenzhen, P.R. China, 518107.  \protect\\
E-mail: {(dx.icandoit@gmail.com, dongx55@mails2.sysu.edu.cn); zhanxlin@mail2.sysu.edu.cn;  yunchao.wei@bjtu.edu.cn; cswei@scu.edu.cn; wangyw@pcl.ac.cn;ymlml@zju.edu.cn; caoxch5@mail.sysu.edu.cn;xdliang328@gmail.com}

}
\thanks{* indicates equal first author. Xiaodan Liang is the Corresponding Author.}
}
%
%

\markboth{IEEE TRANSACTIONS ON XXX
,~Vol.~, No.~xx, xx~2022}%
{Shell \MakeLowercase{\textit{et al.}}: Bare Demo of IEEEtran.cls for Computer Society Journals}
%



\IEEEtitleabstractindextext{%
\begin{abstract}
Our goal in this research is to study a more realistic environment in which we can conduct weakly-supervised multi-modal instance-level product retrieval for fine-grained product  categories. We first contribute the Product1M datasets, and define two real practical instance-level retrieval tasks to enable the evaluations on the price comparison and personalized recommendations. 
For both instance-level tasks, how to accurately pinpoint the product target mentioned in the visual-linguistic data and effectively decrease the influence of irrelevant contents is quite challenging.
To address this, we exploit to train a more effective cross-modal pertaining model which is adaptively capable of incorporating key concept information from the multi-modal data, by using an entity graph whose node and edge respectively denote the entity and the similarity relation between entities.
Specifically, a novel Entity-Graph Enhanced Cross-Modal Pretraining (EGE-CMP) model is proposed for instance-level commodity retrieval, that explicitly injects entity knowledge in both node-based and subgraph-based ways into the multi-modal networks via a self-supervised hybrid-stream transformer, which could reduce the confusion between different object contents, thereby effectively guiding the network to focus on entities with real semantic.  
Experimental results well verify the efficacy and generalizability of our EGE-CMP, outperforming several SOTA cross-modal baselines  like~CLIP~\cite{radford2021learning}, UNITER~\cite{chen2020uniter} and CAPTURE~\cite{CAPTURE}.

\end{abstract}

}

\maketitle

\IEEEdisplaynontitleabstractindextext

%
\IEEEpeerreviewmaketitle

\IEEEraisesectionheading{\section{Introduction}\label{sec:intro}}



\begin{figure}[h]
    \centering
    \includegraphics[width=0.5\textwidth]{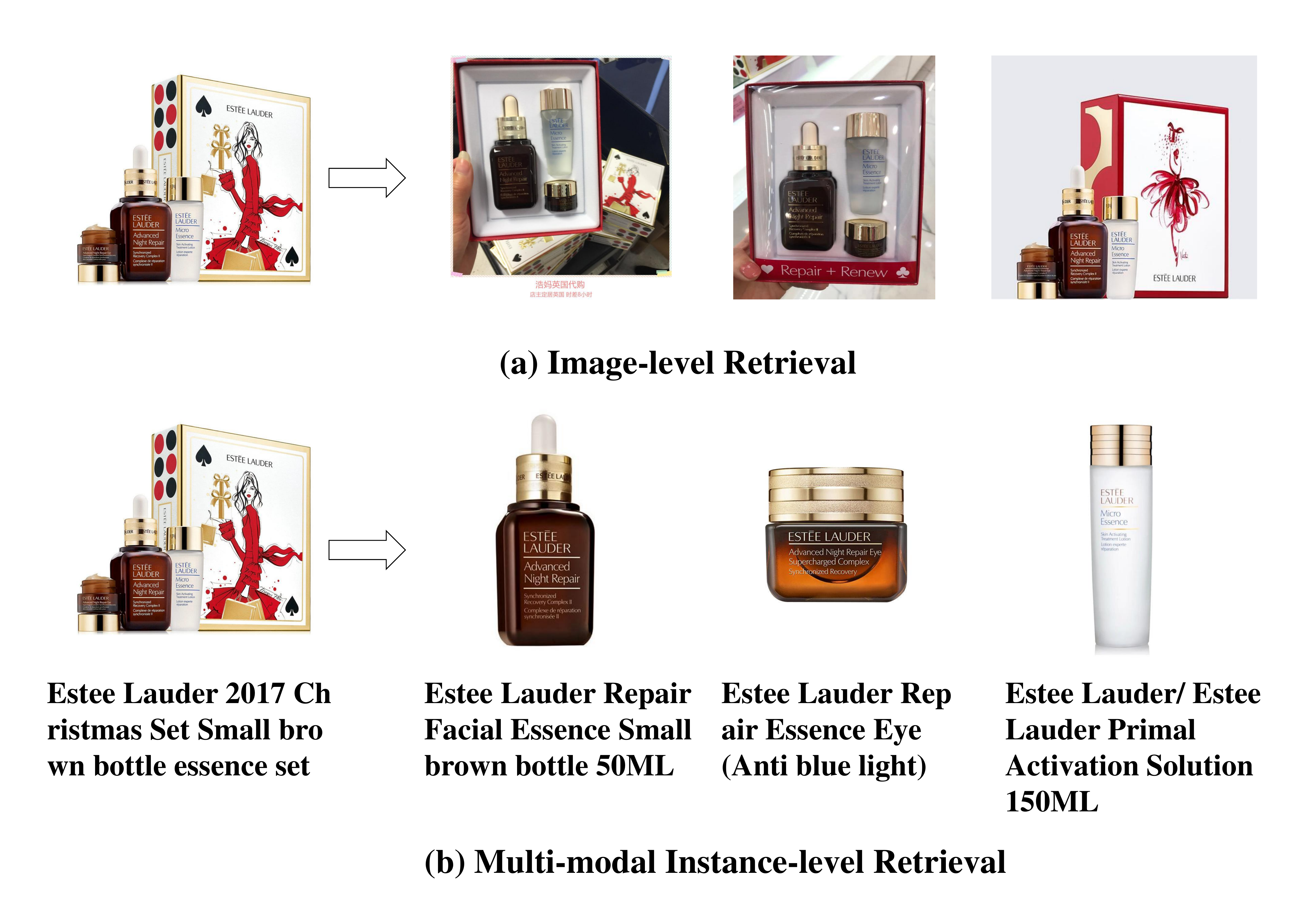}
    \caption{
    The task we propose aims to retrieve \textbf{instance-level} products in \textbf{multi-modal} data.}
    \label{fig_task_difference}
\end{figure}

\IEEEPARstart{E}{-}commerce application as one of the largest multi-modal downstream scenarios has facilitated person's lives. 
And there are many product-based tasks in the application scenario, such as similar item retrieval~\cite{sim_item}, online commodity recommendation~\cite{rahayu2017systematic} and identical product match for cross-price comparison~\cite{vogler2016cancer}. 
With a wide variety of downstream E-commerce applications in the real world, we focus solely on the pretraining of multi-modality data of products. 
In the general scenarios, multi-modal vision-language pre-training model~\cite{lu2019vilbert,Su2020VL-BERT:,li2019visualbert,CAPTURE,chen2020uniter,radford2021learning,tan2019lxmert,BLIP,GLIP} such as CLIP~\cite{radford2021learning} and VilBert~\cite{lu2019vilbert} have shown superior performances on the diverse downstream tasks, such as zero-shot classification~\cite{zero}, cross-modal retrieval~\cite{retrieval} and open-world detection~\cite{detection}. 
They learn visual and textual embedding features from a huge number of image-text pairs acquired from different sources and have quite generalization capability and robustness. 
The fundamental methodology behind these models is either contrastive semantic alignment of images and texts or specific network architectures designed for better common space learning. 
However, these models mentioned above focus more on the attention regions adaptively learned by themselves and are easily influenced by the contents that are not related to the visual objects from each modality, leading to ignoring the importance of concept knowledge, \textit{i.e.}, words or phrases with a certain meaning. 
As an illustration in Figure~\ref{fig_task_difference}, we expect the trained multi-modal network to pay more attention to four typical objects  (\textbf{Fair Water}, \textbf{Facial Cream}, \textbf{Clearner} and \textbf{Clear Lotion})  and  to ignore extra content (\textbf{Four piece Set}) in a more fine-grained application setting. 
If the multi-modal network is trained directly by visual-language contrastive learning, the learned attention region may be \textbf{SK-II}, \textbf{Cream Cleaner}, \textbf{Lotion Four-piece}, 
which is  unsuitable and incorrect to achieve the expected objects of attention. 
In the past few years, there has been increasing attention in designing the knowledge injection~\cite{RoBERTa,oLMpics,KBERT,ERICA} for natural language processing (NLP) tasks to demonstrate and enchance the attention effectiveness of NLP models.
These models are more based on a well fully annotated knowledge graph, which is time-cost and laborious for a new task.
In this paper, we balance the trade-off between knowledge base construction and knowledge extraction, and propose a simple but effective entity-graph enhanced visual-language pretraining model, which explicitly inject the concept knowledge generated from the caption data into the multi-modal model via a special knowledge graph (entity-graph).



Next, We first investigate a realistic problem: \textit{how to perform instance-level~\footnote{Instance-level product retrieval refers to the retrieval of all single products that existed in a product portfolio image.} fine-grained product retrieval given the weakly annotated multi-modality data on a vast scale?} 
In Figure~\ref{fig_task_difference}, we give a comparison of different retrieval patterns to better show the challenge. 
From the figure, we could find that image-level retrieval tends to match retrieval objects from all aspects since it is not capable of discriminating between distinct instances, but multi-modal instance-level retrieval is much more capable of searching for different types of commodities among multi multimodality data. 
Although the challenge has generality and practical importance, it is under-examined enough owing to a lack of high-quality datasets fitting the real scenarios and a clear description of the issue. 
According to relevant studies of product retrieval, uni-modal  \cite{qi2016sketch,bai2018optimization,qayyum2017medical,nurmi2008product} and cross-modal retrieval \cite{wang2017adversarial,feng2014cross,wei2016cross,cao2016deep,wang2015joint,deng2018triplet} take as input unimodal information, \textit{e.g.}, a visual feature or a textual description, and performs a matching search between individual samples. 
Unfortunately, such retrieval strategies severely limit their applicability in many scenarios where both the queries and the targets include multi-modal information. 
More importantly, past researches concentrate on comparatively straightforward scenario, \textit{i.e.}, image-level \footnote{Image-level product retrieval refers to recognizing a specific product instance in a single-product image.} retrieval for single-product images \cite{kuang2019fashion,gu2018multi}, leaving the instance-level aspect of retrieval untouched.

\begin{figure*}[h]
    \centering
    \includegraphics[width=1\textwidth]{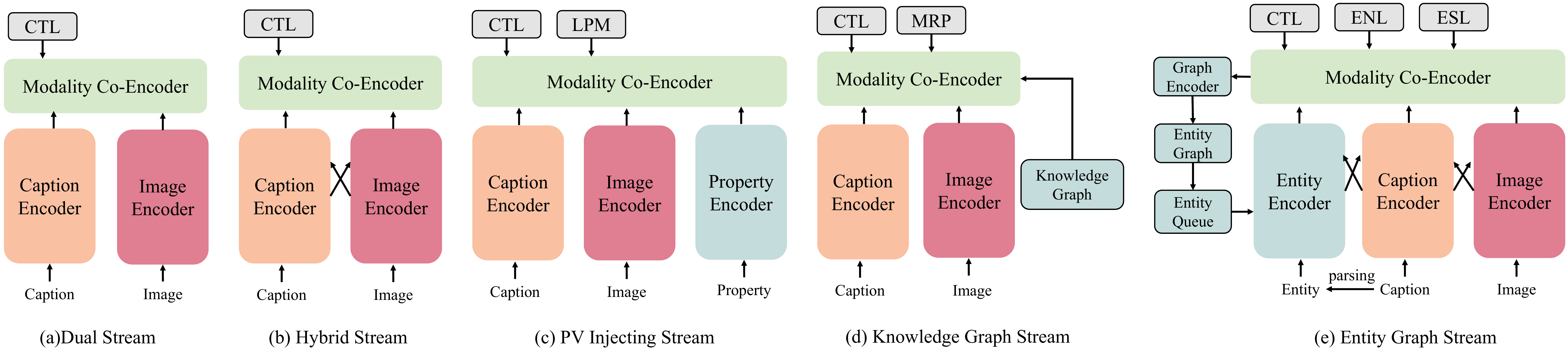}
    \caption{Five categories of visual-language pretraining models. The two left are  the typical network structures. The three right are the knowledge injecting versions. Compared with (c) and (d), our structure (e) builds the entity graph directly from the caption without additional annotation information. In the figure, CTL, LPM, MRP, ENL, ESL denote the ContrasTive Learning, Link Prediction Modeling, Masked Relation Prediction, Entity Node Loss and Entity Subgraph Loss, respectively.}
    \label{fig:pipline}
\end{figure*}

In the paper, we propose a large-scale \textbf{E}ntity-\textbf{G}raph \textbf{E}nhanced \textbf{C}ross-\textbf{M}odal \textbf{P}retraining framework (EGE-CMP) to address these challenges. 
Inspired by the effectiveness of knowledge injection~\cite{RoBERTa,oLMpics,ERNIE,ERNIE3,KEPLER} into large-scale pretrained models for single modality, we model the cross-modal semantic alignment under the supervision of an entity-graph instead of directly using image and text interaction, which  makes the trained model more focused on meaningful entity objects and thus reduce the influence of irrelevant content.
Specifically, we extract the noun entity directly from the caption data via named entity recognition (NER)  tools and obtain the image proposals from the image data via the bottom-up-attention feature representation model~\cite{bottom_up}.   
Then, EGE-CMP is presented to capture the potential synergy of images, entities and title descriptions via entity-graph learning and some task-agnostic masking training. 
We showcase that some widely used cross-modal pretraining methods \cite{lu2019vilbert,li2019visualbert,chen2020uniter,tan2019lxmert} might be ineffective under the multi-instance setting without external knowledge guidance.
In contrast, EGE-CMP utilizes a hybrid-stream architecture that encodes data of different modalities separately and fuses them in  a unified manner, which has been shown to be advantageous for our suggested purposed tasks.
Additionally, we introduce the cross-modal contrastive loss to compel EGE-CMP to align images, texts, and entities, so avoid the mismatch problem introduced by the unsuitable pretext task.
Besides that, an effective entity-graph enhanced module is adopted to enhance the learning capacity for the whole transformer model by enforcing the model to pay more attention to some entities with real semantic information. 
 In our entity graph, the node is each entity data and the edge is generated via the relation between each entity.
To encourage the text-encoder to be aware of actual key entities and improve the understanding ability of the transformer model, node-based and subgraph-based ranking losses are also introduced for the model training, requiring it to accurately distinguish the actual positive samples of entities according to the global entity ranking list. 
As a result, our EGE-CMP model can focus more on the phrasesor words with real semantic meaning, and the text-image correlation can be aligned better. 
 Please note that compared with existing knowledge-enhanced pretraining methods based on the knowledge graph, our EGE-CMP model directly utilizes the knowledge directly extracted from the original caption data without any time-consuming and labor-intensive annotated data.

Furthermore, to bridge this gap and enhance the relevant research, we construct a large-scale dataset from the popular shopping website, named Product1M, proposed for the performance verification of multi-modal instance-level retrieval in the real-world scenarios. 
Product1M has about one million image-caption pairings and is divided into two sorts of samples, \textit{i.e.}, single- and multi-product samples.
The dataset is one of the biggest multimodal datasets available and is the first to be created exclusively for multimodal instance-level retrieval in real-world settings.
Two real-world  tasks, multi-product retrieval and identical-product retrieval, are defined in section~\ref{sec:sec_task_definition}  and used for the model comparisons. 
 In summary, as a large benchmark dataset, our Product1M with ground-truth annotations generated by several human workers such that the performance of the evaluated multi-product and identical-product retrieval algorithms can be better analyzed.

The results of experiments on both multi-product retrieval and identical-product retrieval tasks show the superiority of our EGE-CMP over the SOTA cross-modal baselines, such as ViLBERT~\cite{lu2019vilbert}, CLIP~\cite{radford2021learning}, UNITER~\cite{chen2020uniter}, CAPTURE~\cite{CAPTURE} and so on, on all major criteria by a large margin.
Moreover, extensive ablation experiments are conducted to demonstrate the generalizability of EGE-CMP and investigate various essential factors of our proposed task. 
Finally, the quantitative and qualitative results intuitively prove the entity-graph constructed by caption itself can enhance the representation capability of the transformer models.

The main contributions of our paper are threefold.
\begin{itemize}
\item We conduct Product1M, one of the largest multi-modal product datasets for the real-world instance-level retrieval task, and contribute a subset extracted from Product1M for identical product retrieval for price comparison. 
Different from the existing E-commerce datasets, our Product1M dataset has real-world testing circumstances and abundant categories, which can help advance future research in the high-level difficulties. 
\item We devise a novel Entity-Graph Enhanced Cross-Modal Pretraining (EGE-CMP) \footnote{https://github.com/Xiaodongsuper/Entity-Graph-Enhanced-Cross-Modal-Pretraining-for-Instance-level-Product-Retrieval} framework to learn the instance-level feature representations via injecting the entity knowledge with real semantic information to the visual-text alignment.  We find our entity-graph significantly benefits the text-image cross-modal alignment.
\item Experimental results on both our proposed instance-level retrieval tasks, \textit{i.e.}, multi-product retrieval and same product search demonstrate that the model representation can be effectively augmented by the proposed entity graph. 
Furthermore, both the quantitative and visualized research results intuitively prove the efficacy of our EGE-CMP framework.
\end{itemize}

The study builds on our prior conference presentation, CAPTURE~\cite{CAPTURE}. 
In the version, we add the following features to the conference version: (1) To be suggested, a new goal, identical-product retrieval, must be adequately specified. (2) Additionally, a novel Cross-Modal Pretraining framework called EGE-CMP is presented to integrate entity information into visual-linguistic alignment. 
We demonstrate the superiority of the suggested model EGE-CMP over many state-of-the-art approaches.
 Additionally, we show feature embeddings, do ablation investigations, and perform generalization analysis.

The rest of the paper is divided into the following sections: Section~\ref{sec:related_work} discusses related works. 
Section~\ref{sec:instance_prod} describes the definition of multi-product and identical product retrieval tasks and then introduces our model, including its network architectures and formulation. Section~\ref{sec:exp} exhibits and explains the experimental findings. 
Finally, we make a conclusion in Section~\ref{sec:conclu}.
 
\begin{figure*}
    \centering
    \includegraphics[width=1.0\textwidth]{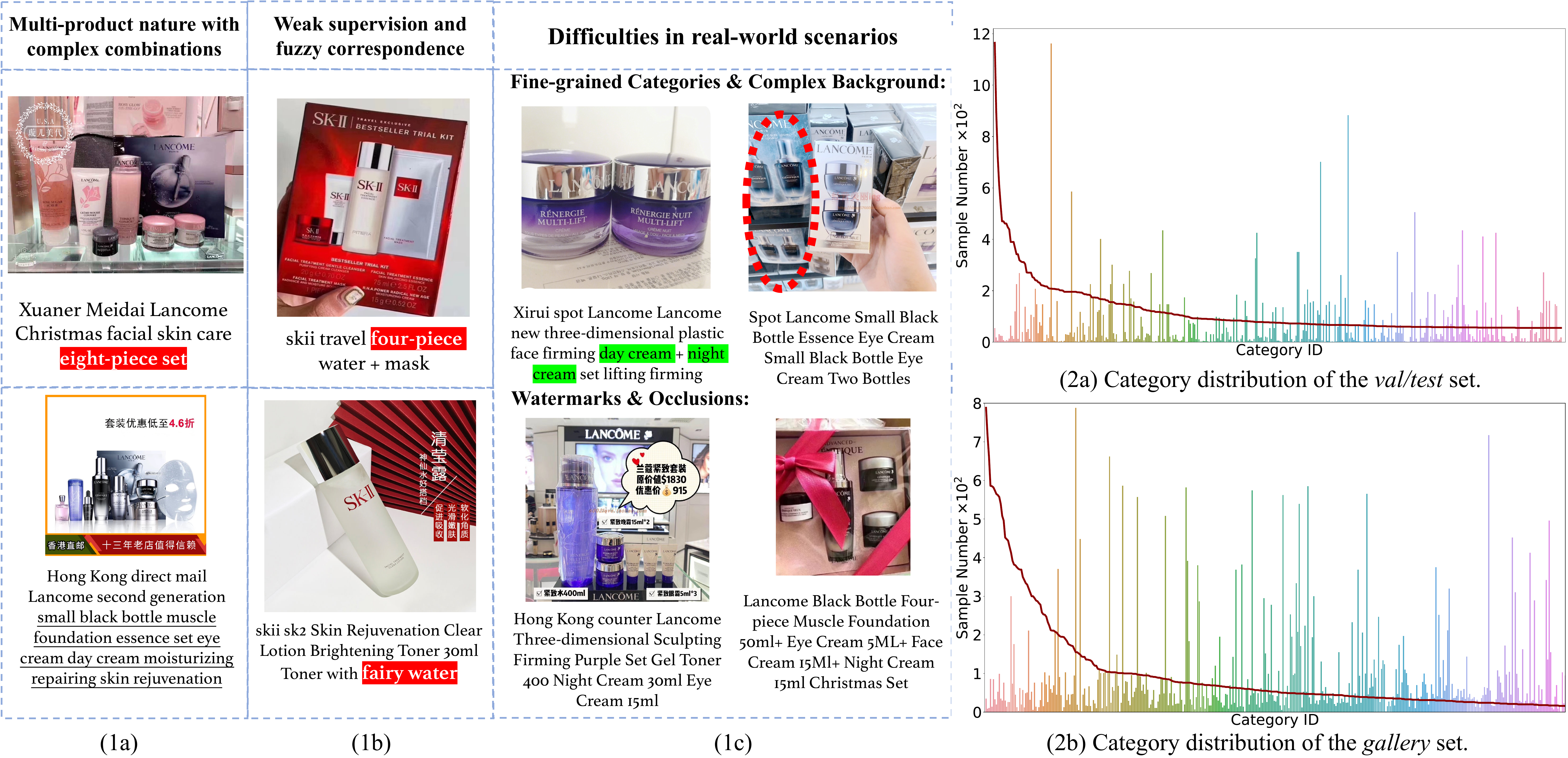}
    \caption{Characteristics and statistics of Product1M: (1a) Complex combinations of single products; (1b) Weak supervision and fuzzy correspondence; (1c) Difficulties in real-world settings; 
    (2) Long-tailed category distribution of Product1M. The line represents the sample size of each category in decreasing order. Product1M comprises a diverse set of categories and the long-tailed class distribution corresponds to real-world cases.}
    \label{fig_challenge_statistics}
\end{figure*}

\section{Related Work}\label{sec:related_work}
\subsection{Visual-Language Datasets.}
According to the collection source, visual-language pre-training datasets can be categorized into at least two main groups:
\textit{1) General Sources.}:
There are several visual-language datasets~\cite{hu2017twitter100k,VG,Flickr,SBU,NUSWIDE,Krapac2010ImprovingWI} collected from the social media platforms (\textit{e.g.} Twitter and Facebook). 
\textbf{Twitter100k} \cite{hu2017twitter100k} is presented for cross-media retrieval with a low degree of supervision.
It is a collection of 100k image-text pairings crawled from the Twitter website. 
There are no limitations on the categories of images, and users supply the textual description in a colloquial language.
Similarly, this dataset emphasizes image-level retrieval and employs single-modal input as the query, which is incompatible with instance-level multi-modal retrieval.  
\textbf{INRIA-Websearch} \cite{Krapac2010ImprovingWI} is a cross-modal retrieval dataset that contains 71,478 image-text pairs connected with 353 unique search queries, such as actors, films, and so on.
    The images are gathered via internet searches, and the textual description is derived from the text that surrounds the images on websites.
In this dataset, cross-modal retrieval methods are used to tackle the text-to-image search issue.
    \textit{2) Product Sources.} For the E-commerce datasets, RPC checkout~\cite{wei2019rpc} and Dress Retrieval~\cite{corbiere2017leveraging} are crawled from the shopping website as two widely used typical datasets. 
\textbf{RPC} \cite{wei2019rpc} comprises 53,739 photos of individual products and 30,000 photographs of checkout.
Three independent levels of annotation are available: category information, point-level annotations, and bounding boxes. 
There is, however, a gap between RPC and real-world settings, since RPC was gathered in a controlled environment and this dataset is devoid of textual information.
\textbf{Dress Retrieval} proposed in \cite{corbiere2017leveraging} collects obtains noisy labeled data scraped from catalogs of E-commerce websites.

As indicated before, our Product1M dataset is an instance-level dataset. Unlike the previous datasets, each picture in the Product1M dataset is connected with a subset of relevant attributes that have been filtered using specified terminology. 
Typically, each picture comprises a single instance, such as a model wearing a garment, with a relatively uncluttered backdrop. 
In summary, our Product1M is significantly different from the aforementioned datasets in three ways: (1) Product1M is the first dataset designed specifically for multi-product and identical product instance-level retrieval tasks, both of which have significant promise in the e-commerce business.
(2) Our Product1M enables multi-modal retrieval in addition to conventional intra- and cross-modal retrieval, when both the query and the target include pictures and text. 
This is a more realistic environment, given the pervasiveness of multimodal information in real-world scenarios. 
(3) The weakly annotated examples of Product1M compel the model to uncover relevant features in the absence of clear labels, endowing the model with a high capacity for generalization to large-scale and noisy data.

\subsection{Intra- and Cross-Modal Retrieval.} 

Intra-modal retrieval~\cite{qi2016sketch,bai2018optimization} has been intensively investigated in the contexts of keyword-based web document retrieval~\cite{ensan2017document} and content-based image retrieval~ \cite{noh2017large}. 
 On the other hand, cross-modal retrieval\cite{wang2017adversarial,feng2014cross,wei2016cross,cao2016deep,wang2015joint,deng2018triplet}  emerges as a prospective route for efficiently indexing and searching vast amounts of data with multiple modalities, and is extensively utilized in search engines~\cite{buttcher2016information,harman2019information} and E-commerce~\cite{ji2017cross,corbiere2017leveraging}, to mention a few. 
 However, these techniques~\cite{nurmi2008product,lin2018regional,corbiere2017leveraging,wei2019rpc,wang2016effective} are often limited to unimodal inputs, making them inapplicable to a large number of real-world circumstances non which both the queries and targets include multi-modal information.

\subsection{WSOD: Weakly Supervised Object Detection.} 
By learning from cheaper or publicly accessible data, WSOD~\cite{tang2018pcl,2018Generative,2019Cap2Det} decreases its over-reliance on fine-grained labeling.
PCL~\cite{tang2018pcl} builds proposal clusters iteratively to aid in the construction of instance classifiers.
Pseudo labels elicited from picture labels~\cite{2018Generative} and unstructured textual descriptions such as captions~\cite{2019Cap2Det} are also useful for improving WSOD performance.
WSOD, on the other hand, often relies on a fixed-size collection of predefined classes and is thus unsuitable for our proposed task, since class labels are not available and categories may be changed dynamically.

\subsection{Visual-Linguistic Pretraining.}

The development of the vision-language pretraining model~\cite{Oscar,fashionbert,Unimo,Violet,TACo,flip,vilt,add_Florence,add_M6,add_SimVLM,add_slip,add_SUVVL,add_unip,add_wenlan} has advanced tremendously, with joint representations often constructed using a multimodal fusion transformer network topology.
Existing pre-trained models for vision-language often learn image-text semantic alignment using a multi-layer Transformer architecture, such as Bert~\cite{Bert}, on multi-modal input in a shared cross-modal space.
They may be loosely divided into two categories based on their network structural differences:
1) Single-stream models~\cite{li2019visualbert,Su2020VL-BERT:,chen2020uniter,Unimo,Oscar,fashionbert,vilt} in a unified architecture encode the integrated multi-modal characteristics. 2) Two-stream models~\cite{lu2019vilbert,tan2019lxmert,Violet,TACo,flip,add_Florence,add_wenlan,add_unip,add_SUVVL}, on the other hand, use distinct encoders for inputs with varying modalities.
These approaches are not optimized for instance-level retrieval, and we demonstrate how they may be defective as a result of network architecture design flaws and incorrect pretext tasks.
Additionally, since real-world semantic information cannot be effectively learned during the network training phase, performance increases are restricted.

\subsection{Knowledge-Enhanced Modeling.} In the language models, the knowledge-based model aims to incorporate knowledge, concepts and relation into the pretrained models~\cite{Bert,RoBERTa}, which proved to be beneficial to language understanding~\cite{oLMpics}. 
The existing methods can be coarsely broadly classified into two types: implicit knowledge modeling and explicit knowledge injection. 
The former attempts on implicit knowledge modeling usually consist of entity-level masked modeling~\cite{ERNIE,KBERT,ERICA}, entity-based replacement prediction~\cite{PENCYC}, knowledge embedding loss as regularization~\cite{KEPLER} and universal knowledge-text prediction~\cite{ERNIE3,KLMo}. 
In contrast to implicit knowledge modeling, the latter separately maintains a group of parameters for representing structural knowledge. Such methods~\cite{ERNIE,KBERT,PENCYC,ERNIE3,KLMo,ERICA} usually require a heterogeneous information fusion component to fuse multiple types of features obtained from the text and knowledge graphs. 
In our setting, it is very hard to build a knowledge graph with good annotations from scratch to fit the multi-modal data. 
In the paper, with the purpose of knowledge modeling and visual-language pretraining, and improving the representation capability of visual-text models, we explore the solutions for knowledge enhanced cross-modal pretraining, where knowledge is represented as an entity parsed from the original caption data and the relationship between different pieces of knowledge are learned via node-based and subgraph-based ranking losses. 

Figure~\ref{fig:pipline} shows five typical vision-and-language models. According to whether external knowledge is utilized, these models can be generally fall into the following categories: 1) models without knowledge supervision that fall under Figure~\ref{fig:pipline} (a) and (b), 2) knowledge-based models that fall under Figure~\ref{fig:pipline} (c), (d) and (e). Our proposed model EGE-CMP belongs to the second type. Meanwhile, different from the existing models as seen in Figure~\ref{fig:pipline} (c) and (d), our model directly use the entity knowledge extracted from the caption rather than human labeling, (\textit{e.g.} property and knowledge graph) which extremely save the labor and calculation cost. Our proposed EGE-CMP is the first model of type Figure~\ref{fig:pipline} (e) where the entity knowledge drawn from the caption data is explicitly injected into the model via graph learning.

\section{Instance-Level Retrieval on Product1M}
\label{sec:instance_prod}
In this section, we begin by describing the two instance-level tasks, followed by our proposed EGE-CMP model. 
The task definition is given in Section~\ref{sec:sec_task_definition}. The dataset analysis is presented in Section~\ref{subsec:dataset}. 
The details of our proposed Entity-Graph Enhanced Cross-Modal Pretraining (EGE-CMP), including model architecture, masked relation prediction, cross-modal contrastive learning, and the model inference, are elaborated in Section~\ref{subsec:method}.

\subsection{Task Definition}
\label{sec:sec_task_definition}

\subsubsection{Multi-Product Retrieval}
A product sample $(I, C)$ is an image-text pair where $I$ is the product image and $C$ is the caption.
    Given the \textit{gallery} set of single-product samples $\mathcal{S}=\{\mathcal{S}_i |\mathcal{S}_i= (I_\mathcal{S}^i,C_\mathcal{S}^i)\}$ and the set of multi-product samples $\mathcal{P}=\{\mathcal{P}_i | \mathcal{P}_i=(I_\mathcal{P}^i,C_\mathcal{P}^i)\}$, the task is to retrieve and rank the single-products that appear in the query sample $\mathcal{P}_i$, \textit{i.e.}, to predict a list $RETR^i = [id^i_1, id^i_2, \cdots, id^i_k, \cdots, id^i_N]~~ \forall \mathcal{P}_i \in \mathcal{P}$, where $id^i_k$ corresponds to a specific single-product sample in $\mathcal{S}$.
    
\subsubsection{Identical-Product Retrieval} 
    In the task, we do not distinguish between single-products and multi-product. 
    Similar to the multi-product retrieval task, given the \textit{gallery} set of product samples $\mathcal{S}=\{\mathcal{S}_i |\mathcal{S}_i= (I_\mathcal{S}^i,C_\mathcal{S}^i)\}$ and the set of product samples $\mathcal{P}=\{\mathcal{P}_i | \mathcal{P}_i=(I_\mathcal{P}^i,C_\mathcal{P}^i)\}$, the task is to retrieve and rank the products that appear in the query sample $\mathcal{P}_i$, \textit{i.e.}, to predict a list $RETR^i = [id^i_1, id^i_2, \cdots, id^i_k, \cdots, id^i_N]~~ \forall \mathcal{P}_i \in \mathcal{P}$, where $id^i_k$ corresponds to a same product sample with differnet angles, views, colors and backgrounds in $\mathcal{S}$. The task focuses on the higher level of fine-grained match compared with traditional fine-grained retrieval tasks.

\subsection{Dataset Analysis}\label{subsec:dataset}
\subsubsection{Dataset Statistics}
\label{sec:sec_statistics}

We gather several product samples from E-commerce websites for 49 different brands.
Following that, these image-text samples are sorted manually into single- and multi-product categories based on the product information associated with them.

\begin{table*}
\centering
\begin{tabular}{c|cccc|cccc}
\toprule[1pt]
{{Dataset}}& {\footnotesize{}{\#samples}} &  {\footnotesize{}{\#categories}} &  {\footnotesize{}{\#instances}}
&  {\footnotesize{}{\#obj/img}} 
&  {\footnotesize{}{weak supervision}}
&  {\footnotesize{}{multi-modal}}
&  {\footnotesize{}{instance-level retrieval}}
&  {\footnotesize{}{multi-task}}
\tabularnewline
\midrule[1pt]
Twitter100k~\cite{hu2017twitter100k} & 100,000 & - & - & - & \checkmark &\checkmark & - & -\tabularnewline
Visual Genome~\cite{VG} &  108,000 & -  & - & - &\checkmark &\checkmark &-  & -\tabularnewline
Flickr30K~\cite{Flickr} & 31,000 & -  & - & - &\checkmark &\checkmark &-  &-\tabularnewline
SBU~\cite{SBU} &  890,000 & -  & - & - &\checkmark &\checkmark &- &- \tabularnewline
NUS-WIDE~\cite{NUSWIDE}  & 269,648 & 81  & - & - &\checkmark &\checkmark &- &- \tabularnewline
INRIA-Websearch~\cite{Krapac2010ImprovingWI} &  71,478 & 353  & - & - &\checkmark &\checkmark &- & -\tabularnewline
\midrule[1pt]
RPC checkout~\cite{wei2019rpc} & 30,000 & 200 & 367,935 & 12.26 &-  &- &- &- \tabularnewline
Dress Retrieval~\cite{corbiere2017leveraging} & 20,200 & - & $\sim$20,200 & $\sim$1.0&\checkmark &\checkmark &- &- \tabularnewline
\textbf{Product1M(Ours)} & \textbf{1,182,083}  & \textbf{458} & \textit{92,200} & \textit{2.83} & \checkmark & \checkmark  &\checkmark & \checkmark \tabularnewline

\bottomrule[1pt]
\end{tabular}
\caption{Comparisons between different datasets. `-' indicates inapplicable. The \#instances and \#obj/img of Product1M are in italics since there are no instance labels for the \textit{train} set and we only count the instances in the \textit{val} and \textit{test} set. Product1M is one of the largest multi-modal datasets and the first to be specifically designed for instance-level retrieval in real-world applications.}
\label{tab_statistics}
\end{table*}


 For \textbf{the multi-product retrieval problem}, Product1M is divided into the \textit{train}, \textit{val}, \textit{test}, and gallery sets.
The \textit{train} set has 1,132,830 samples, including both single- and multi-product samples, while the \textit{val} and \textit{test} sets contain only multi-product examples, totaling 2,673 and 6,547 samples, respectively.
The \textit{gallery} set contains 40,033 single-product samples in 458 categories, 392 of which occur in the \textit{val} and \textit{test} sets, while the others serve as interference objects for validating the robustness of a retrieval technique.
The \textit{gallery}, \textit{val}, and \textit{test} sets are annotated with class labels for evaluation purposes only; they are not used in the training process, and the \textit{train} set samples are not annotated.
Table~\ref{tab_statistics} and Figure~\ref{fig_challenge_statistics}.  summarize statistics of the data for Product1M.

\begin{figure}
    \centering
    \includegraphics[width=0.5\textwidth]{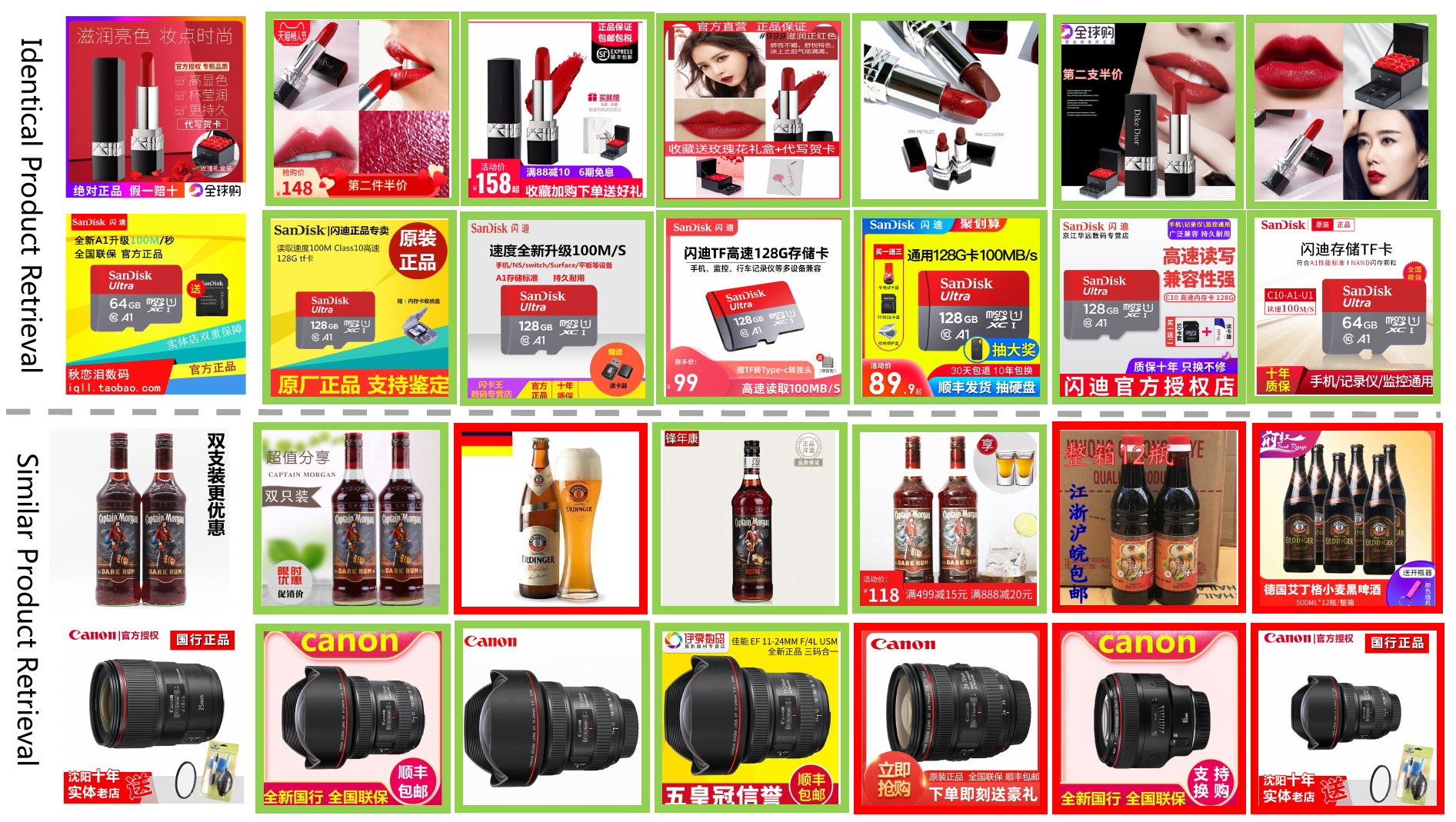}
    \caption{Several comparison demos of identical-prouduct retrieval and  similarity product retrieval. Accurate/Inaccurate matched results are framed in green/red boxes.}
    \label{fig:identical_product_retrieval}
\end{figure}

For the \textbf{identical-product retrieval task}, we extract a subset from the our Product1M to construct \textit{train}, \textit{test} and \textit{gallery} sets. The wholse subset totally contains 208,228 samples with 819 unique products  including mixed single-product and multi-product samples. The \textit{train}, \textit{gallery} and \textit{test} set respectively contain 192,164, 14,628 and 1,436 samples. All test samples are annotated by at least 5 crowdsourced workers, and each worker label a pair of data, one is query data and another is gallery data, to determine whether both samples are the same products. As depicted in Figure~\ref{fig:identical_product_retrieval}, identical product retrieval is more fine-grained compared with the traditional similarity retrieval task.

\subsubsection{Dataset Characteristics}
\label{sec:sec_challenge}

\noindent \textbf{Multi-product nature and complex combinations:} Multi-product photos are prevalent on E-commerce websites and serve as the query images for product retrieval at the instance level. 

As seen in Figure~\ref{fig_challenge_statistics}(1a), products may be structured in a variety of ways and layouts, with a high number of instances. 
Due to the abundance and variety of fine-grained single-product samples, complex combinations appear in various portfolio photos.

\noindent \textbf{Weak supervision and fuzzy correspondence:}
    We investigate retrieving data in two common modalities, namely pictures and text.
    In comparison to comparable datasets with precise class names, the oversight provided by commodity captions is inadequate and often uninformative.
    In Figure~\ref{fig_challenge_statistics}(1b), we illustrate many sorts of difficult samples.
    Various samples' captions include abbreviations, \textit{i.e.}, a reduced version of several goods.
    However, an acronym such as 'eight-piece set' has no information about the items.
    The second sort of sample contains irrelevant data, with the commodities specified in the title appearing in the picture but not in the title, or vice versa.
    The vast dispersion of imprecise connection between pictures and titles complicates instance-level retrieval even more.

\noindent \textbf{Consistency with real-world scenarios:}
    In Figure~\ref{fig_challenge_statistics}(1c), we provide several difficult examples.
    They may feature a complicated backdrop with irrelevant elements, amorphous watermarks, or substantial clutter that obscures the product information.
    Certain items from various categories may seem practically identical save for the packaging description, \textit{e.g.}, \textit{day cream} versus \textit{night cream}.
    As seen in Figure~\ref{fig_challenge_statistics}(2a,2b),  the long-tailed distribution of Product1M fits nicely with real-world settings.

\begin{figure*}
    \centering
    \includegraphics[width=1.0\textwidth]{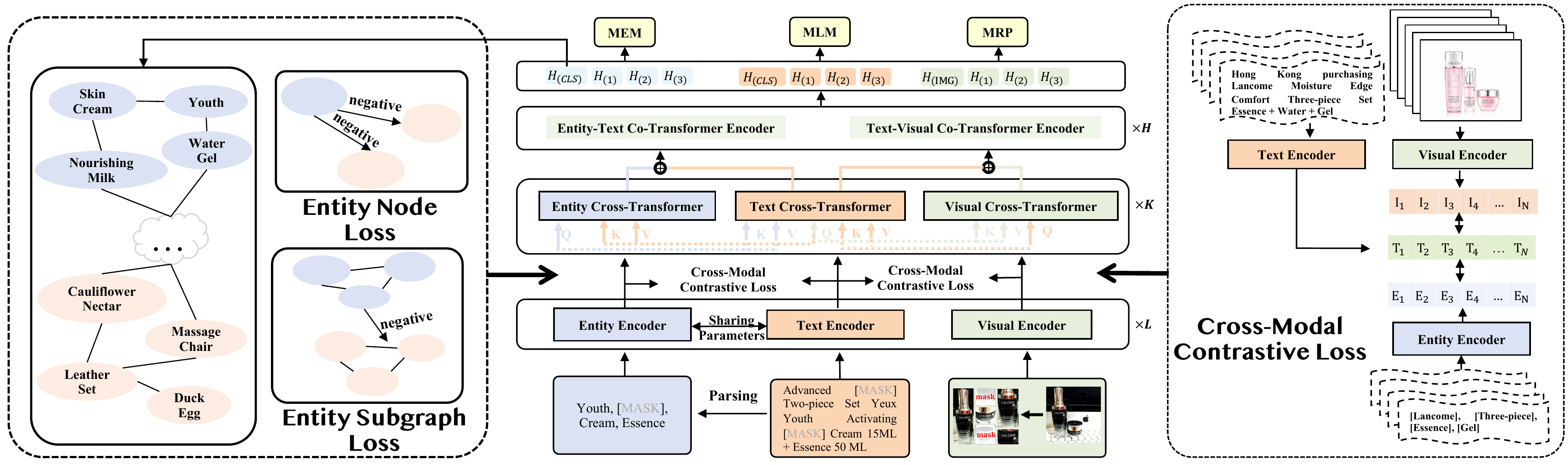}
    \caption{An overview of our Entity-Graph Enhanced Cross-Modal Pretraining (EGE-CMP) model. First,  three kinds of data are fed into the EGE-CMP model, where entity data is parsed by NER tools (middle-bottom).  
    Then, we employ a cross-modal contrastive loss to force modality data to come closer with their positive samples and away from the negative samples (right).
    After cross-transformer and CO-transformer, we utilize the AdaGAE~\cite{AdaGAE} graph network to construct an entity-graph and propose node-based and subgraph-based ranking losses for making the multi-modal network pay more attention to the meaningful entity and reducing the ambiguity between different entities (left). }
    \label{fig_overall_EGE-CMP}
\end{figure*}
\subsection{Methodology}\label{subsec:method}

Figure~\ref{fig_overall_EGE-CMP} depicts the general architecture of our end-to-end visual-language model (EGE-CMP). In the network architecture, EGE-CMP contains three kinds of transformer modules: single-modal transformer (SMT), cross-modal transformer (CMT) and co-modal transformer (COMT) from bottom to top. 
Meanwhile, according to the function modules, our model consists of three modules: Masked Relation Learning, Cross-modal Contrastive Learning and Entity-graph Enhanced Learning. 
We begin this section by elaborating on the architectural design of 
EGE-CMP in Section \ref{sec_EGE-CMP_arch} including detailed functions of three kinds of transformer modules and the direction of data flow. 
Then we describe three main function modules that enables the self-supervised learning of EGE-CMP in Section \ref{sec_EGE-CMP_masked}, Section \ref{sec_EGE-CMP_contrastive} and Section~\ref{sec_entity}.  
Finally, in Section~\ref{sec_inference}, we show the inference procedure for instance-level retrieval.

\subsubsection{Architectural Design of EGE-CMP}
\label{sec_EGE-CMP_arch}
As mentioned above, our EGE-CMP contains SMT, CMT and COMT three kinds of transformer structures. 
EGE-CMP takes an image-text pair $(I, C)$ as input where $I$ is the product image and $C$ is the caption data. Image $I$ are first processed by bottom-up-attention model~\cite{bottom_up} to obtain the proposal features $p=\{p_1, p_2,..,p_m\}$ where objects are most likely to be detected and their corresponding position information $q=\{q_1,q_2,..,q_m\}$. 
Caption $C$ are parsed by NER analysis tools to obtain the noun entities $E_{C}=\{E_1, E_2,...\}$. 
Based on the above preparation, SMT extracts the embeddings $v, t, e$ for three different inputs (image, caption and entity). 
Then CMT performs cross-modal fusion respectively on $v$ and $t$ by exchanging key-value pairs to generate the new representations $h_v$ and $h_{t_1}$. 
Similarly,  we utilize the same operation to produce the new representations $h_e$ and $h_{t_2}$. 
To keep the representation consistent, we take the average of $h_{t_1}$ and $h_{t_2}$ to obtain the new representations of text $h_t$. 
Finally, {$h_v$, $h_t$} and {$h_e$, $h_t$} are separately concatenated in the feature dimension and fed into two COMTs to produce joint image-text-entity representations $f$ for multi-product retrieval and identical-product retrieval tasks.
We give the detailed explanations of each transformer structure in below.

\noindent\textbf{Single-Modal Transformer (SMT).} For a image input $I$, the position information $q$ are aligned with the same dimension as the proposed feature $p$ using a linear projection layer and added with $p$ as the input to obtain the preliminary image embeddings $u_v$:
\begin{equation}
\label{eq:v}
    u_v=\mathrm{SMT}(p+\mathrm{LinearProj}(q))
\end{equation}

For a caption input $C$ or entity input $E$, it is processed by WordPiece and tokenized into word tokens ${w_i}_{i=1}^{L}$, where $L$ is the length of tokens in caption $C$ or entity $E$.
Our SMT embeds the discrete word token $w_i$ into a continue word embedding $u_t$ and $u_e$:
\begin{equation}
\begin{aligned}
    &u_t=\mathrm{SMT}(w_t)  \\
    &u_e=\mathrm{SMT}(w_e)
\end{aligned}
\label{eq:te}
\end{equation}
where $w_t$ and $w_e$ are word tokens of caption input $C$ and entity input $E$.

\noindent\textbf{Cross-Modal Transformer (CMT).}
By exchanging key-value pairs in the multi-headed attention mechanism, the Cross-Transformer is utilized to construct the inter-modal relations between the two distinct modalities.
 We separately learn the new embeddings $h_{t_1}$ and $h_{v}$ between image and text modalities. With the same way, the new embeddings $h_{t_2}$ and $h_{e}$ are generated by CMT. Due to the downstream tasks only accessing image and caption data, we select the $h_{t_1}$ as the output $h_t$ to serve the COMT module.
The detailed process can be shown as following.
 \begin{equation}
\begin{aligned}
    &h_{t_1}, h_v = \mathrm{CMT}(u_t, u_v)  \\
    &h_{t_2}, h_e = \mathrm{CMT}(u_t, u_e) \\
    &h_{t} = (h_{t_1} + h_{t_2})/2
\end{aligned}
\label{eq:CMT}
\end{equation}

\noindent\textbf{CO-Modal Transformer (COMT).} Given image features $h_v$, caption featuress $h_t$ and concatenated entity features $h_e$, COMT performs cross-modal fusion over all $\{h_{v_i}\}_{i=1}^{T}$, $\{h_{t_i}\}_{i=1}^{L}$ and $\{h_{e_i}\}_{i=1}^{L}$  for joint visual-language-entity learning. 
Because the caption and entity data could express the semantic information of different levels, we construct two branches co-transformer to solve the problem. 
One branch is to learn the common representation between $h_v$ and $h_t$. 
We add a common learnable position embedding $q_{vt}$ to caption features $h_t$ and image features $h_v$, to incorporate sequence ordering and learn the low-level semantic information. Then, the image and caption representations after position embedding $q_{vt}$ and following a sepical $[CLS]$ token are concatenated as the input to the branch of COMT. Similarly, we also use the same operation to process $h_t$ and $h_e$. After that, two group fused presentations $[f_v, f_{t_1}]$ and $[f_e, f_{t_2}]$ are obtained. To learn the fused text representation, $f_{t_1}$ and $f_{t_2}$ are concatenated and adopted a linear projection layer to generate the final output $f_t$. The joint features $f$ are learned as:

\begin{equation}
\begin{aligned}
    &f_v, f_{t_1} = \mathrm{COMT_1}([[\mathrm{CLS}], [u_t, u_v]+q_{vt}])  \\
    &f_e, f_{t_2} = \mathrm{COMT_2}([[\mathrm{CLS}], [u_t, u_e]+q_{et}])  \\
    &f_t = \mathrm{LinearProj}([f_{t_1}, f_{t_2}]) \\
    &f = [f_v, f_t, f_e]
\end{aligned}
\label{eq:COMT}
\end{equation}

\subsubsection{EGE-CMP by Masked Multi-Modal Learning}
\label{sec_EGE-CMP_masked}
We utilize several pretext tasks to enable the self-supervised learning of EGE-CMP and benefit from large-scale multi-modal data. 
For modality-wise feature learning, we adopt three masked multi-modal modeling tasks, \textit{i.e.}, Masked Language Modeling task (MLM), Masked Region Prediction task (MRP) and Masked Entity Modeling (MEM) following the standard BERT \cite{Bert} and VisualBERT \cite{li2019visualbert}. 
Masked Language Modeling reconstructs the masked token information to improve language understanding with the aid of visual and knowledge representations. Masked Region Prediction recovers the masked visual features to enhance the visual object perception. Masked Entity Modeling improves the semantic knowledge understanding, empowering models to make sense of common concepts with real significance.

\noindent \textbf{Masked Language Modeling (MLM).} In the task,  approximately 15\% of texts and are masked out and the remaining inputs are used to reconstruct the masked tokens $x$ from the joint features $h_t$ learned from COMT module.  The masked tokens $h_{t}^{m}$ are adopted by a fully-connected (FC) layer ($FC_{MLM}$) to be projected to the discrete pre-defined word space for classification:
\begin{equation}
\begin{aligned}
x_{i}^{\prime} &=\mathrm{FC}_{\mathrm{MLM}}\left(h_{t}^{\mathrm{m}}\right), \\
\mathcal{L}_{\mathrm{MLM}} &=-\mathbb{E}\left[\frac{1}{\left|\mathcal{M}^{\mathrm{MLM}}\right|} \sum_{i \in \mathcal{M}^{\mathrm{MLM}}} \log P\left(x_{i} \mid x_{i}^{\prime}\right)\right],
\end{aligned}
\end{equation}

where  $\mathcal{M}^{\mathrm{MLM}}$ is the index set of masked text tokens.

\noindent \textbf{Masked Region Prediction (MRP).} Similar to the MLM, we randomly mask out proposal inputs $p_{v}^{m}$ with approximately 15\%. Different from the discrete work token, in the task, we directly regress the masked features $p_i$, which is supervised by the features extracted by the bottom-up-attention model~\cite{bottom_up} with an MSELoss. The masked proposal inputs $h_t^m$ are fed into a fully-connected (FC) layer  ($FC_{MRM}$) and project to a binary space for determining whether the input is masked :

\begin{equation}
\begin{aligned}
p_{i}^{\prime} &=\mathrm{FC}_{\mathrm{MRM}}\left(p_{v}^{\mathrm{m}}\right), \\
\mathcal{L}_{\mathrm{MRM}} &=\frac{1}{2\left|\mathcal{M}^{\mathrm{MRM}}\right|} \sum_{i \in \mathcal{M}^{\mathrm{MRM}}} \left|p_{i}^{\prime} - p_{i}^{t} \right|^2,
\end{aligned}
\end{equation}
where $\mathcal{M}^{\mathrm{MRM}}$ is the index set of masked proposal inputs, and $p_i$ and $p_i^t$ are the predicted results and ground-truth.

\noindent \textbf{Masked Entity Modeling (MEM).} In MEM, about 15\% of entity tokens are masked out and the remaining entity inputs are used to reconstruct the masked entity $x_e$ from the joint features $h_e$ learned from COMT module.  The masked tokens $h_{e}^{m}$ are adopted by a fully-connected (FC) layer ($FC_{MEM}$) to be projected to the discrete pre-defined word space for classification:
\begin{equation}
\begin{aligned}
x_{e_i}^{\prime} &=\mathrm{FC}_{\mathrm{MEM}}\left(h_{e}^{\mathrm{m}}\right), \\
\mathcal{L}_{\mathrm{MEM}} &=-\mathbb{E}\left[\frac{1}{\left|\mathcal{M}^{\mathrm{MEM}}\right|} \sum_{i \in \mathcal{M}^{\mathrm{MEM}}} \log P\left(x_{e_i} \mid x_{e}^{\prime}\right)\right],
\end{aligned}
\end{equation}

where $\mathcal{M}^{\mathrm{MEM}}$ is the index set of masked entity tokens.

\subsubsection{EGE-CMP by Cross-Modal Contrastive Loss}
\label{sec_EGE-CMP_contrastive}
Apart from intra-modal feature learning, EGE-CMP is enable to construct coherent representations of multi-modal inputs and learn their alignment. 
To do this, we leverage inter-modality contrastive learning~\cite{chen2020simple,radford2021learning} to align the visual and textual modalities.
There are $2N$ data points in total for a minibatch of $N$ image-text samples.
We consider the related image-text pairings to be $N$ positive pairs and the remaining $2(N-1)$ mismatched pairs to be negative pairs.
Formally, given an image-text pair $(x_i,x_j)$ and their encoded features $(\tilde{x}_i, \tilde{x}_j)$, the cross-modal contrastive loss for this positive pair is computed as:
\begin{equation}
   \mathcal{L}_\mathrm{VT}(x_i, x_j)=-\log \frac{\exp \left(\operatorname{sim}\left(\tilde{x}_i, \tilde{x}_j\right) / \tau\right)}{\sum_{k=1}^{2 N} \mathbbm{1}_{[k \neq i]} \exp \left(\operatorname{sim}\left(\tilde{x}_i, \tilde{x}_k\right) / \tau\right)},
\end{equation}
where $\operatorname{sim}(\boldsymbol{u}, \boldsymbol{v})=\boldsymbol{u}^{\top} \boldsymbol{v} /\|\boldsymbol{u}\|\|\boldsymbol{v}\|$ computes the cosine similarity of $(\boldsymbol{u}, \boldsymbol{v})$ pairs, $\tau$ denotes the temperature parameter, $\mathbbm{1}_{[k\ne i]}$ is a binary indicator function that returns 1 iff $k \ne i$.

This kind of contrastive loss promotes similarity between the encoded characteristics of positive pairings from various modalities while differentiating between the encoded properties of negative pairs.

To make the text encoder implicitly injected with abundant semantic information. Single entity data are stacked as the input to be aligned with corresponding caption data. The cross-modal constrastive learning between   a pair of entity-caption data $(e_i, x_j)$ is represented as:
\begin{equation}
   \mathcal{L}_{\mathrm{ET}}(e_i, x_j)=-\log \frac{\exp \left(\operatorname{sim}\left(\tilde{e}_i, \tilde{x}_j\right) / \tau\right)}{\sum_{k=1}^{2 N} \mathbbm{1}_{[k \neq i]} \exp \left(\operatorname{sim}\left(\tilde{e}_i, \tilde{x}_k\right) / \tau\right)},
\end{equation}

\subsubsection{EGE-CMP by Entity-Graph Enhanced Learning} 
\label{sec_entity}

\begin{figure}
    \centering
    \includegraphics[width=0.5\textwidth]{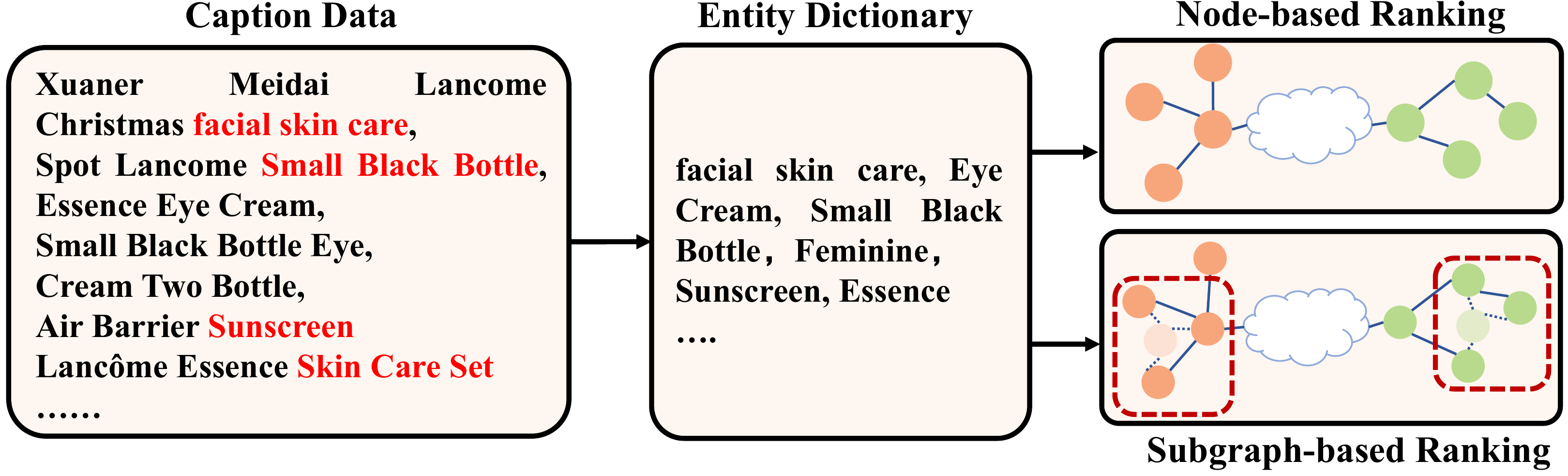}
    \caption{An overview of the EGE module. The module has three steps: 1) Entity object extraction, 2) Entity dictionary construction, 3) Entity relation cultivation: Node-based and Subgraph-based.}
    \label{fig:our_EGE_module}
\end{figure}

To improve the generalization capability and entity discrimination of our framework,  we develop an effective entity-graph enhanced (EGE) module, which can explicitly inject the entity knowledge into the model training process.
Given a caption $C$ with a set of entities $E_C=\{E_1, E_2,...\}$ directly extracted from caption data themselves, EGE aims to learn a semantic graph to connect each entity data for more accurate instance-level retrieval.
In the EGE module, a unique entity queue $l$ is first 
initialized. 
Then each entity in the queue is encoded by our COMT module to obtain the joint embeddings $f_{e_i}$, and fed to the AdaGAE network~\cite{AdaGAE} that is an effective graph clustering network for learning better graph semantic relationships $f_{g_i}$ between each other. 
Finally, we respectively propose node-based ranking loss and subgraph-based ranking loss to make each entity pay more attention to its semantically close neighbors. 
The process of the module is shown in  Figure~\ref{fig:our_EGE_module}.

Inspired by the memory bank of contrastive learning, we also keep a queue $l$ in our training process for obtaining more accurate negative sample selection.  The queue is initialized with an entity dictionary that is the set of all entity data. 
Hence the length of the queue is equal to the number of unique entities of our dataset.  

To update the queue $l$, we feed all entities into the COMT module to obtain the feature embeddings $l_f$. 
Then based on the embeddings $l_f$, we utilize AdaGAE~\cite{AdaGAE} to learn the graph embedding $l_g$ by integrating relationships with different nodes for better node representations. 
In our training process, the cosine similarity $S_g$ of $l_g$ is to determine which entity is negative sample. 
In the netwok training process, for a minibatch of data, we perform negative sample selection from this small batch of data based on similarity $S_g$. For any entity $h_{e_{i}}$ in a minibatch of data with $l_m$ length, we randomly select one $h_{e_{k}}$ of the last $k$ entity sample as the negative sample. The node-based ranking loss is represented as:

\begin{equation}
\label{eq:node}
\mathcal{L}_\mathrm{node}(h_{e_i}, h_{e_k}, y)=\mathcal{\max} (0,-y *(h_{e_i}-h_{e_k})+\mathrm{margin})
\end{equation}
where $y$ equal to 1 and margin is to set 0. 

Meanwhile, we also propose the sub-graph based loss for keeping the relationships of nearest neighbor graphs for both positive and negative data points. In our paper, the feature embedding of any sub-graph is represented with a global-add pooling via aggregating information of their $k$ nearest neighbor node. For the positive and negative  nearest neighbor subgraph $h_{g_i}$ and $h_{g_k}$ for the node $h_{e_i}$, we have the following formulation 

\begin{equation}
\label{eq:subgraph}
\mathcal{L}_\mathrm{subgraph}(h_{g_i}, h_{g_k}, y)=\mathcal{\max} (0,-y *(h_{g_i}-h_{g_k})+\mathrm{margin})
\end{equation}
where $y$ equal to 1 and margin is to set 0.

\subsubsection{Inference for Instance-Level Retrieval}
\label{sec_inference}
For both the single- and multi-product samples, the proposal-level features are extracted via the bottom-up-attention network~\cite{bottom_up} pre-trained on the Visual Genome dataset~\cite{VG}, the captions are used as input and the entity is to set the same values to EGE-CMP.
The COMT layer produces $f_{v}$ and $f_{t}$ as the feature embeddings of the visual and textual inputs, respectively, during inference. 
Then, $f_{v}$ and $f_{t}$ are concatenated and normalized to get the instance's joint representations.
Additionally, since the Text/Visual and Text/Entity Transformers are supervised using cross-modal contrastive loss, we could believe that concatenating the feature embeddings of COMTs for retrieval is useful.
Our retrieval model then uses the generated features as input.
After generating the cosine similarity matrix between an instance and the $gallery$ set samples,  the single-product samples with the greatest similarity are selected and returned for each query for both multi-product and identical-product retrieval tasks.

\section{Experiments}
\label{sec:exp}

\subsection{Implementation Details}
\label{sec_implement}
\begin{figure}[t!]
    \centering
    \includegraphics[width=0.5\textwidth]{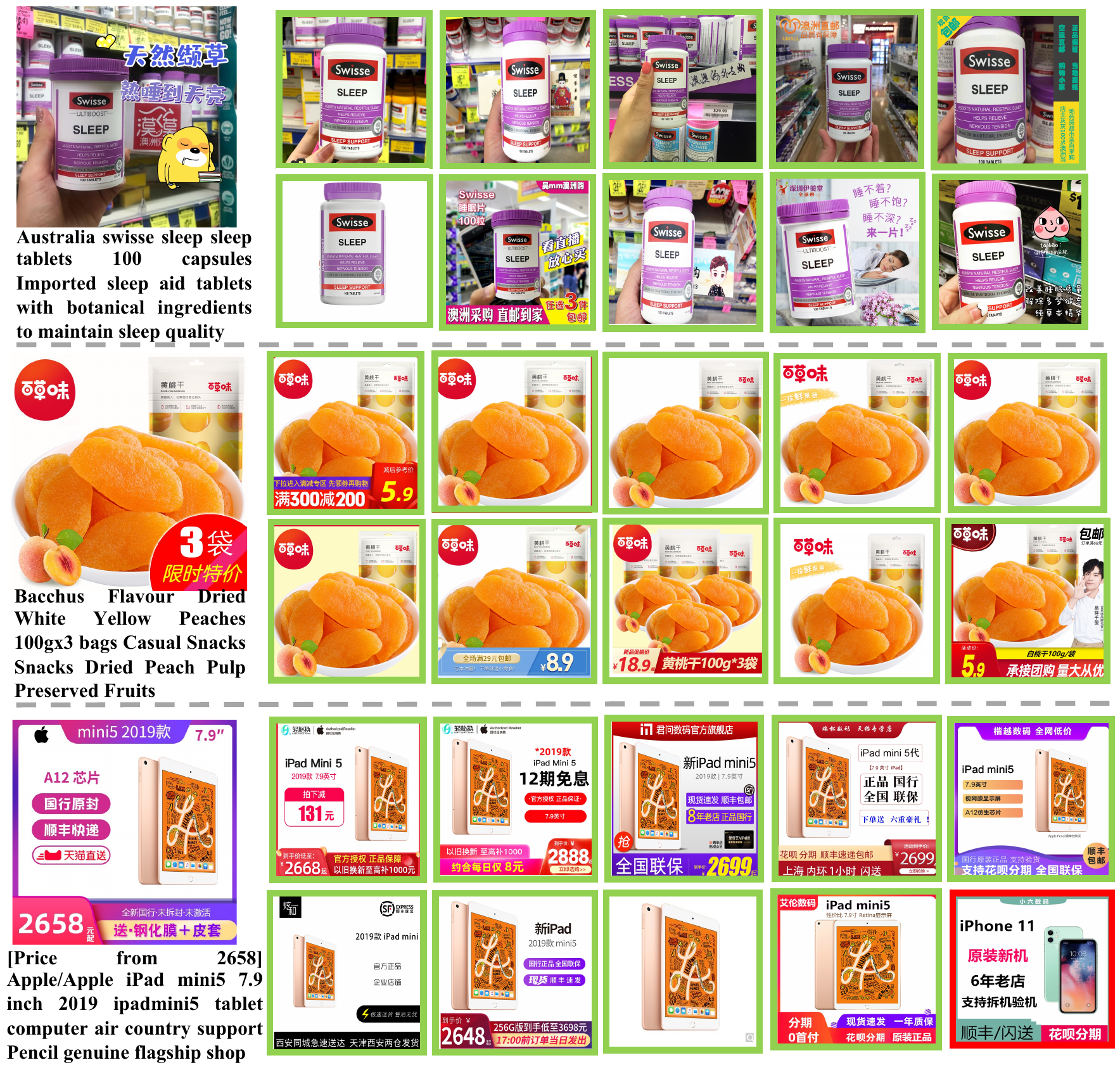}
    \caption{More retrieval visualizations of EGE-CMP on the identical product retrieval task. On the left are the query images. Green boxes indicate correct retrieval images, whereas red boxes indicate faulty retrieval images. The captions of the retrieved  samples have been omitted for the sake of simplicity.}
    \label{fig:fig_retrieval_results}
\end{figure}

We directly utilize the bottom-up-attention model pre-trained~\cite{bottom_up} on the VG dataset~\cite{VG} to extract proposal-wise features for the visual data and employ the BERT~\cite{Bert} model to initialize the linguistic transformer of our EGE-CMP. 
We limit the number of input areas to between 10 and 36 by picking regions with predicted confidence greater than a predefined threshold as in \cite{lu2019vilbert}. 
The region's features are then flattened and input into EGE-CMP using RoIAlign.
The number of the Text/Visual/Entity Transformer, Text/Visual and Text/entity Cross-Transformer, and Co-Transformer is set to $L=4, K=4,$ and $H=4$, respectively, for a total of 12 transformer layers.
We set the hidden state size of EGE-CMP and other baselines to 768 for a fair comparison.
The transformer blocks in EGE-CMP have hidden state size of 768 with 8 attention heads.
Meanwhile, we separately attach a 512 dimensional fully connected layer after Co-Transformer and Text/Visual/Entity Transformer for task-agnostic masked learning and cross-modal contrastive learning. 
By  concatenating the features from the final Co-transformer layer, a 1024 dimensional feature vector is obtained for retrieval, which is also the same for other baselines.
The maximum sequence length for the sentence is set to 36. 
Our EGE-CMP model is trained with a total batch size of 128 for 10 epochs on 8 RTX 3090 GPUs. 
And an Adam \cite{kingma2014adam} optimizer is used with an initial learning rate of 1e-4 and a linear learning rate decay schedule is adopted in the EGE-CMP training process. 
Temperature parameter $\tau$ is set to 0.07.
The entity numbers are limited to 12,000 for faster training speed.
EGE-CMP generates instance features from texts and proposal-wise features during inference.
Before retrieval, the instance features of the image/text are all normalized and then concatenated as the overall representations. 
During retrieval, we compute cosine similarities between each box query with single-product samples using the overall representations mentioned above and score all retrieval single-product results based on their similarity. 
The top-$N$ samples are returned as the retrieval results.
Fig~\ref{fig:fig_retrieval_results} shows more retrieval results by EGE-CMP.
To provide a fair comparison with other baselines, we conduct all experiments using the same training settings and assessment procedures.


\begin{table*}[h]
\centering
\resizebox{\textwidth}{!}{
\begin{tabular}{l|ccc|ccc|ccc}
\toprule[1pt]
Method& {\footnotesize{}{mAP@10}} & {\footnotesize{}{mAP@50}} & {\footnotesize{}{mAP@100}} & {\footnotesize{}{mAR@10}} & {\footnotesize{}{mAR@50}} & {\footnotesize{}{mAR@100}} & {\footnotesize{}{Prec@10}} & {\footnotesize{}{Prec@50}} & {\footnotesize{}{Prec@100}} \tabularnewline
\midrule[1pt]
\textit{Image-based} & 47.46  &  39.79  & 37.06   & 20.81  & 16.12  & 14.63  & 36.28  & 31.20  & 29.20  \\
\textit{Text-based} & 68.85 &  62.21  & 60.56   & 24.01  & 17.48  & 16.44  & 62.48  & 59.26  & 54.86  \\
ViLBERT \cite{lu2019vilbert} & 77.54 &  74.73  &  73.74  &  31.31 & 22.69  & 23.48  &  75.42 & 72.03  &  70.85 \\
LXMERT \cite{tan2019lxmert} & 78.62  & 76.57   &  75.29 & 31.63  & 25.98  &  24.92 &75.57  &  74.29 & 72.81  \\ 
CLIP* \cite{radford2021learning}  & 76.03  & 75.33   &  73.66  &  30.90  &  25.59  & 24.59 & 73.07  & 72.42 &70.03 \\
VL-BERT \cite{Su2020VL-BERT:} &76.20   & 75.22   & 72.84   &30.76 & 25.87  &24.97  &  73.26 & 72.41  &70.83  \\ 
VisualBERT\cite{li2019visualbert} &77.68 & 76.33   & 75.00  & 31.89  & 26.66 & 25.43  &75.47  & 73.09 & 69.35 \\
UNITER \cite{chen2020uniter} &79.00  & 77.37   &  75.73  & 32.05 &26.40  &  25.22 & 76.39  &74.42 & 73.99  \\
CAPTURE~\cite{CAPTURE} & 81.49 & 79.38 & 78.81 & 32.99 &  26.86 &\textbf{26.46} &77.27 & 76.54 &\textbf{75.00}   \\
\textbf{EGE-CMP (Ours)} & \textbf{86.91} & \textbf{82.68} & \textbf{80.57} & \textbf{32.99} &  \textbf{26.92} & 25.68 & \textbf{82.59} & \textbf{77.89} & 73.72   
\tabularnewline
\bottomrule[1pt]
\end{tabular}}
\caption{Comparison of multi-product retrieval performance using various intra- and cross-modal self-supervised baselines.}
\label{tab:instance_pro_retrieval}
\end{table*}

\begin{table*}[h]
\centering
\resizebox{1.5\columnwidth}{!}{
\begin{tabular}{l|ccc|ccc}
\toprule[1pt]
Method& {\footnotesize{}{mAP@10}} & {\footnotesize{}{mAP@50}} & {\footnotesize{}{mAP@100}}  & {\footnotesize{}{Prec@10}} & {\footnotesize{}{Prec@50}} & {\footnotesize{}{Prec@100}} \tabularnewline
\midrule[1pt]
\textit{Image-based} &52.90   &44.65    & 41.51     &52.90   &35.22   &33.75  \\
\textit{Text-based} &70.21 &66.21    &63.88     &70.21   &63.87   &60.54   \\
ViLBERT \cite{lu2019vilbert} &81.17   &83.45     &81.88   &81.74  &63.07   &56.99  \\
LXMERT \cite{tan2019lxmert} &81.09   &82.02     & 81.48    &81.09  &64.76  &58.78   \\ 
CLIP* \cite{radford2021learning}  &81.80   &82.90    &81.11    &81.80   &62.43  &57.98  \\
VL-BERT \cite{Su2020VL-BERT:} &81.39   &83.02    &81.44      &81.39   &64.05   &58.41   \\ 
VisualBERT\cite{li2019visualbert} &79.90   &81.91    &80.32    &79.90  &62.39   &55.87   \\
UNITER \cite{chen2020uniter} &81.74   &83.45    &81.88     &81.74   &63.07   &56.99   \\
CAPTURE~\cite{CAPTURE}  &82.03  &83.60  & 82.04  &82.03  &65.22 &59.03   \\
\textbf{EGE-CMP (Ours)} & \textbf{84.66} & \textbf{86.37} & \textbf{84.79}  & \textbf{84.66} & \textbf{67.76} & \textbf{62.00}   \tabularnewline
\bottomrule[1pt]
\end{tabular}}
\caption{Comparison of identical product retrieval using different intra- and cross-modal self-supervised baselines.}
\label{tab:iden_pro_retrieval}
\end{table*}

\subsection{Evaluation Metrics}
We use two commonly used measures~\cite{weyand2020GLDv2,CGW17} Precision ($\mathrm{Prec}@N$), mean Average Precision ($\mathrm{mAP} @ N$) and a new  metric mean Average Recall ($\mathrm{mAR} @ N$) to evaluate the instance-level retrieval performances. 
Because retrieving each and every product extensively is unnecessary and impractical in many cases, we report $\mathrm{mAP}@N$, $\mathrm{mAR}@N$, and $\mathrm{Prec}@N$, where $N = 10, 50, 100$.
$\mathrm{Prec} @ N$ evaluates the average accuracy of the top-$N$ predictions per image and is widely used in the retrieval literature.
$\mathrm{Prec} @ N(q)$ is defined as follows:
\begin{equation}
    \mathrm{Prec}@N(q) = \frac{1}{N}\sum_{i=1}^N \mathrm{acc}_q(i),
\end{equation}
where $\mathrm{acc}_q(i)$ is a binary indicator function that returns $1$ when the $i$-th prediction is correct for the $q$-th query and $0$ otherwise. 
$\mathrm{mAP} @ N$ is computed as the average $\mathrm{AP@}N$ per image, where $\mathrm{AP@}N(q)$ is computed as follows:
\begin{equation}
    \mathrm{AP@} N(q)=\frac{1}{\min \left(m_{q}, N\right)} \sum_{k=1}^{N} \mathrm{P}_{q}(k) \mathrm{rel}_q(k),
\label{eq_ap}
\end{equation}
where $m_q$ is the total number of ground truth images, \textit{i.e.}, corresponding single-product images that appear in the $q$-th query image, $P_q(k)$ is the precision at rank $k$ for the $q$-th query, and $\mathrm{rel}_q(k)$ is a binary indicator function that returns $1$ when the $k$-th prediction is correct for the $q$-th query and $0$ otherwise. 

To evaluate the recall of instance-level retrieval results, we propose metric $\mathrm{mAR} @ N$, which can be computed as the average $\mathrm{AR@}N$ per image. 
$\mathrm{AR@}N(q)$ is re-defined in our paper as follows:
\begin{equation}
    \mathrm{AR@}N(q) = \frac{1}{C_q}\sum_{c=1}^{C}\mathbbm{1}_c(q)\mathrm{min}(1, \frac{RETR^q_c}{\mathrm{min}(\lfloor r_q^c \cdot N \rfloor, G_c)})
\end{equation}
where $C$ is the total number of single-product categories in the \textit{gallery} set, $C_q$ equals to the number of existing categories in the $q$-th query, $\mathbbm{1}_c(q)$ is a binary indicator function that returns 1 when class $c$ exists in the $q$-th query, $RETR^q_c$ is the number of retrieved products belonging to class $c$ for the $q$-th query, $G^c$ is the number of ground truths belonging to class $c$ in the \textit{gallery} set, $r_q^c$ is the instance ratio\footnote{For instance, for a 2A+3B query image, $r_q^A=0.4$ and $r_q^B=0.6$.} of category $c$ in the $q$-th query, and $\lfloor\cdot\rfloor$ is rounding operation. 
As per the equation, $\mathrm{AR@}N(q)$ takes the category distribution into account, \textit{i.e.},
the inclusion of instance ratio is informative for evaluating both the correctness and diversity of a retrieval algorithm and guarantees that some trivial results are not overestimated\footnote{For a 2A+3B query image and $N=100$, $\mathrm{AR@}100(q)$ returns 0.51 and 1.0 for retrieval results 1A+99B and 40A+60B, respectively.}.

\subsection{Weakly-Supervised Instance-Level Retrieval}
\label{sec_main_results}
Table~\ref{tab:instance_pro_retrieval} and Table~\ref{tab:iden_pro_retrieval} show the retrieval performances compared with several intra- and cross-modal models for both multi-product and identical product retrieval tasks.

\begin{table*}[]
\label{tab:loss_ablation}
\centering
\footnotesize
\resizebox{1.7\columnwidth}{!}{
\begin{tabular}{c|cccccc|c}
\toprule[1pt]
Image$_{based}$ &VT-Masked & E-Masked & CTR &Node &Subgraph  & Concat & mAP/mAR/Prec \\
\midrule[1pt]
\#1& $\checkmark$ &  & $\checkmark$ &\   &  &$\checkmark$ & 80.94 / 33.54 / 76.37  \\
\#2&$\checkmark$ & $\checkmark$ & $\checkmark$ && &  $\checkmark$  & 83.28 / 32.19  / 78.72 \\
\#3&  $\checkmark$& $\checkmark$   & $\checkmark$ &$\checkmark$ & & $\checkmark$  & 83.86 / 31.40 / 77.71  \\
\#4& $\checkmark$ & $\checkmark$ &$\checkmark$ &$\checkmark$ & $\checkmark$ & $\checkmark$ & \textbf{84.74} / \textbf{32.91} / \textbf{78.76}\\
\midrule[1pt]
Text$_{based}$ &VT-Masked & E-Masked & CTR &Node &Subgraph  & Concat & mAP/mAR/Prec \\
\midrule[1pt]
\#1& $\checkmark$ &  &$\checkmark$ &\   &  &$\checkmark$ & 78.23 / 31.32 / 74.74  \\
\#2& $\checkmark$& $\checkmark$ & $\checkmark$ && &   $\checkmark$ & 83.48 / 32.39  / 78.34 \\
\#3&  $\checkmark$& $\checkmark$   & $\checkmark$ &$\checkmark$ & &  $\checkmark$ & 83.50 / 32.03 / 78.41  \\
\#4& $\checkmark$& $\checkmark$ &$\checkmark$&$\checkmark$& $\checkmark$ & $\checkmark$ & \textbf{85.25} / \textbf{32.83} / \textbf{80.52}\\
\midrule[1pt]
Image-Text$_{based}$ &VT-Masked & E-Masked & CTR &Node &Subgraph  & Concat & mAP/mAR/Prec \\
\midrule[1pt]
\#1& $\checkmark$  &  & $\checkmark$ &\   &  &  $\checkmark$ & 81.49 / 32.99 / 77.27  \\
\#2& $\checkmark$ & $\checkmark$  & $\checkmark$  && &$\checkmark$     & 85.71 / 33.02  / 81.80 \\
\#3& $\checkmark$  &$\checkmark$   & $\checkmark$  &$\checkmark$ & & $\checkmark$   & 85.76 / 32.60 / 81.91  \\
\#4& $\checkmark$ & $\checkmark$  &$\checkmark$ &$\checkmark$ & $\checkmark$  & $\checkmark$   & \textbf{86.91} / \textbf{32.99} / \textbf{82.59}\\
\toprule[1pt]
\end{tabular}
}
\caption{The impact of different pretext tasks and cross-modal contrastive loss. Evaluation for $N=100$. The term 'VT-Masked' refers to two masked multi-modal pretext tasks, \textit{i.e.}, MLM and MRP. 'E-Masked' denotes the MEM task. 'CTR' refers to cross-modal contrastive loss.}
\label{tab:loss_ablation}
\end{table*}

\subsubsection{Multi-product Retrieval Performance}
\noindent \textbf{Intra-Modal Schemes.} 
Our EGE-CMP model is compared to two intra-modal schemes: \textit{Image-based} and \textit{Text-based} schemes.
To perform image-based retrieval, the Image SMT layer are stacked depicted in Section~\ref{sec_EGE-CMP_arch} and the same image input and pretext tasks are used as same as EGE-CMP, namely MRP.
Similarly, the Text SMT layer is stacked for text-based retrieval and employ just the textual input and MLM pretext task.
Additionally, the depth of these two models is increased to 24 layers in order to maintain the same number of parameters as EGE-CMP.
According to the experimental results shown in Table~\ref{tab:instance_pro_retrieval} and Table~\ref{tab:iden_pro_retrieval},  we can observe that these two schemes trail significantly behind since they are restricted to unimodal data, indicating the importance of modeling the relationship between multimodality data.
In Section~\ref{sec_layer_ablation}, we present further experiment findings to substantiate this conclusion.

\noindent\textbf{Cross-Modal Schemes.} In Table~\ref{tab:instance_pro_retrieval}, we compare EGE-CMP to many commonly used self-supervised cross-modal pretraining approaches, including state-of-art single- and two-stream Vision-language models, as well as a state-of-art zero-shot classification model, \textit{i.e.}, CLIP~\cite{radford2021learning}.
CLIP* refers to a CLIP-like architecture that encodes images and text separately and is trained with a contrastive learning.
Notably, EGE-CMP exceeds all of these benchmarks on the majority of three major retrieval criteria.
Two-stream models, such as LXMERT~\cite{tan2019lxmert}, ViLBERT~\cite{lu2019vilbert}, and CLIP*, do not perform better than single-stream models, indicating that the manner of fusion of cross-modal information is an important component.
Please note that by using entity knowledge, our EGE-CMP outperforms CAPTURE when employing the hybrid transformer architecture.
We attribute EGE-CMP's better performance to its entity knowledge injection. Finally, we examine the influence of various layer types in Section~\ref{sec_layer_ablation}.

\subsubsection{Identical product Retrieval Performance}
As the same setting with multi-product retrieval task,  We also perform the identical-product retrieval on all the compared methods and our EGE-CMP model. 
Table~\ref{tab:iden_pro_retrieval} presents the experimental results.  
These results indicate that performances of all methods have a slight improvement due to the limited number of the gallery set. 
Meanwhile, compared with multi-product retrieval, the performances of the multi-modal pretraining models, \textit{i.e.}, ViLBert~\cite{lu2019vilbert}, UNITER~\cite{chen2020uniter}, CAPTURE~\cite{CAPTURE} and so on,  are much closer, which indicates that identical product retrieval is a more difficult task. 
In both mAP and Prec metrics, our EGE-CMP still obtains the best performance. 
It can effectively prove that the commonsense words or phrases existing in caption data are beneficial to performance improvement.

\subsection{Impact of Entity Pretext Tasks and Graph Ranking Loss}
\label{sec_pretext_ablation}
Table~\ref{tab:loss_ablation} provides the experimental results of our EGE-CMP under different entity loss settings as well as different feature representations, \textit{e.g.} Masked Entity Modeling, Entity Node Loss and Entity Subgraph Loss. 
We apply the Masked Entity Modeling (E-Masked) at the Entity/Text and Text/Visual COMT layer to make the representation more focused on the tokens related to the real contexts, which is beneficial to the modality combination learning.
In mAP metrics, E-Masked Loss helps our model improve the retrieval performance by 2.8, 7.0 and 5.2 respectively (\#1 vs \#2) under the different feature representations, and we find it of certain help when added to the deeper layers. 
Moreover, after making Entity Node Loss act on the features from the Text/Visual Transformer with that from the Co-Transformer for retrieval, it further improves mAP metrics, respectively (\#2 vs \#3). 
We also verify the effectiveness of Entity Subgraph Loss and observe that it is quite useful to further improve the model performance, which suggests that the improvement mainly comes from the Entity learning.

\begin{table}[t!]
\centering
\footnotesize
\resizebox{0.95\columnwidth}{!}{
\begin{tabular}{lcc|c}
\toprule[1pt]
Model & Config & Depth & mAP/mAR/Prec
\tabularnewline
\midrule[1pt]

w/o-Cross & (6,0,6) & 12 & 85.9 / 32.2 / 82.6 \\
w/o-Co & (6,6,0) & 12 & 85.8 / 32.6 / 82.4 \\  
w/o-T/V/E & (0,6,6) & 12 &  50.0 / 18.7 / 35.9  \\
\midrule[0.7pt]
EGE-CMP-A & (2,5,5) & 12 &  85.2 / 31.4 / 79.7   \\
EGE-CMP-B & (5,2,5) & 12 & 86.9 / 33.2 / 82.4  \\
EGE-CMP-C & (5,5,2) & 12 &  86.7 / 33.3 / 82.2  \tabularnewline
\midrule[0.7pt]
EGE-CMP-S & (2,2,2) & 6 & 83.7 / 30.1 / 77.7 \\
EGE-CMP-L & (8,8,8) & 24 &  86.9 / 30.9 / 82.5   \\
\midrule[0.7pt]
\textbf{EGE-CMP} & \textbf{(4,4,4)} & \textbf{12} & \textbf{86.9 / 32.9 / 82.6} \\

\bottomrule[1pt]
\end{tabular}}
\caption{Performance of different layer combinations. 
Evaluation for $N=10$.}
\label{tab:ablation_layer}
\end{table}

\begin{table}
\centering
\footnotesize
\resizebox{\columnwidth}{!}{
\begin{tabular}{c|c|c|c}
\toprule[1pt]
 & Metric & $N=10$ &  $N=50$  \tabularnewline
\midrule[1pt]
\multirow{3}{*}{\rotatebox{90}{\scriptsize 5 brands}} & 
mAP$@N$ &  63.3 / 64.5 / \textbf{70.2} &  60.7 / 62.2 / \textbf{67.4}  \tabularnewline
& mAR$@N$ & 23.2 / 24.7 / \textbf{30.1}  &  19.2 / 20.1  / \textbf{22.9}   \tabularnewline
&Prec$@N$ & 56.5 / 57.1 / \textbf{64.7}  &  56.5 / 57.7 / \textbf{66.3}   \tabularnewline
\midrule[0.7pt]
\multirow{3}{*}{\rotatebox{90}{\scriptsize 10 brands}} &
mAP$@N$ & 56.8 / 58.2 / \textbf{64.7} &  54.1 / 55.2 / \textbf{60.2}   \tabularnewline
&mAR$@N$ & 19.6 / 20.5 / \textbf{25.3}  &  16.4 /17.4 / \textbf{20.4}   \tabularnewline
&Prec$@N$ & 50.2 / 51.9 / \textbf{56.4}  &  51.3 / 52.7 / \textbf{57.5}   \tabularnewline
\midrule[0.7pt]
\multirow{3}{*}{\rotatebox{90}{\scriptsize 20 brands}} 
& mAP$@N$ & 42.6 / 43.3 / \textbf{46.8}  & 36.7 / 37.3 / \textbf{40.2}   \tabularnewline
&mAR$@N$ & 17.5 / 17.6 / \textbf{18.3}  &  12.9 / 13.2 / \textbf{15.0}   \tabularnewline
&Prec$@N$ & 32.2 / 32.4 / \textbf{36.4}  &  32.9 / 33.1 / \textbf{37.5}  \tabularnewline
\bottomrule[1pt]
\end{tabular}}
\caption{Comparison of zero-shot retrieval performance. Listed in the following order: LXMERT/UNITER/EGE-CMP.}
\label{tab:zero_shot}
\end{table}

\begin{table}
\centering
\footnotesize
\resizebox{\columnwidth}{!}{
\begin{tabular}{l|ccc}
\toprule[1pt]
Method& {\footnotesize{}{mAP$@100$}} &  {\footnotesize{}{mAR$@100$}} &  {\footnotesize{}{Prec$@100$}} \tabularnewline
\midrule[1pt]
UNITER-single & 86.56 & 80.82 & 80.82 \tabularnewline
LXMERT-single & 86.05 & 80.59 & 80.59 \tabularnewline
CAPTURE-single & 88.24 & 83.33 & 83.33 \tabularnewline
EGE-CMP-single & \textbf{91.00} & \textbf{85.42} & \textbf{86.37} \tabularnewline
\midrule[0.7pt]

\end{tabular}}
\caption{Ablation study of single-product retrieval. Note that for single-product retrieval, the metric Prec$@N$ is equivalent to mAR$@N$ since an image contains just one category.}
\label{tab:ablation_detector}
\end{table}

\begin{figure}[t!]
    \centering
    \includegraphics[width=0.5\textwidth]{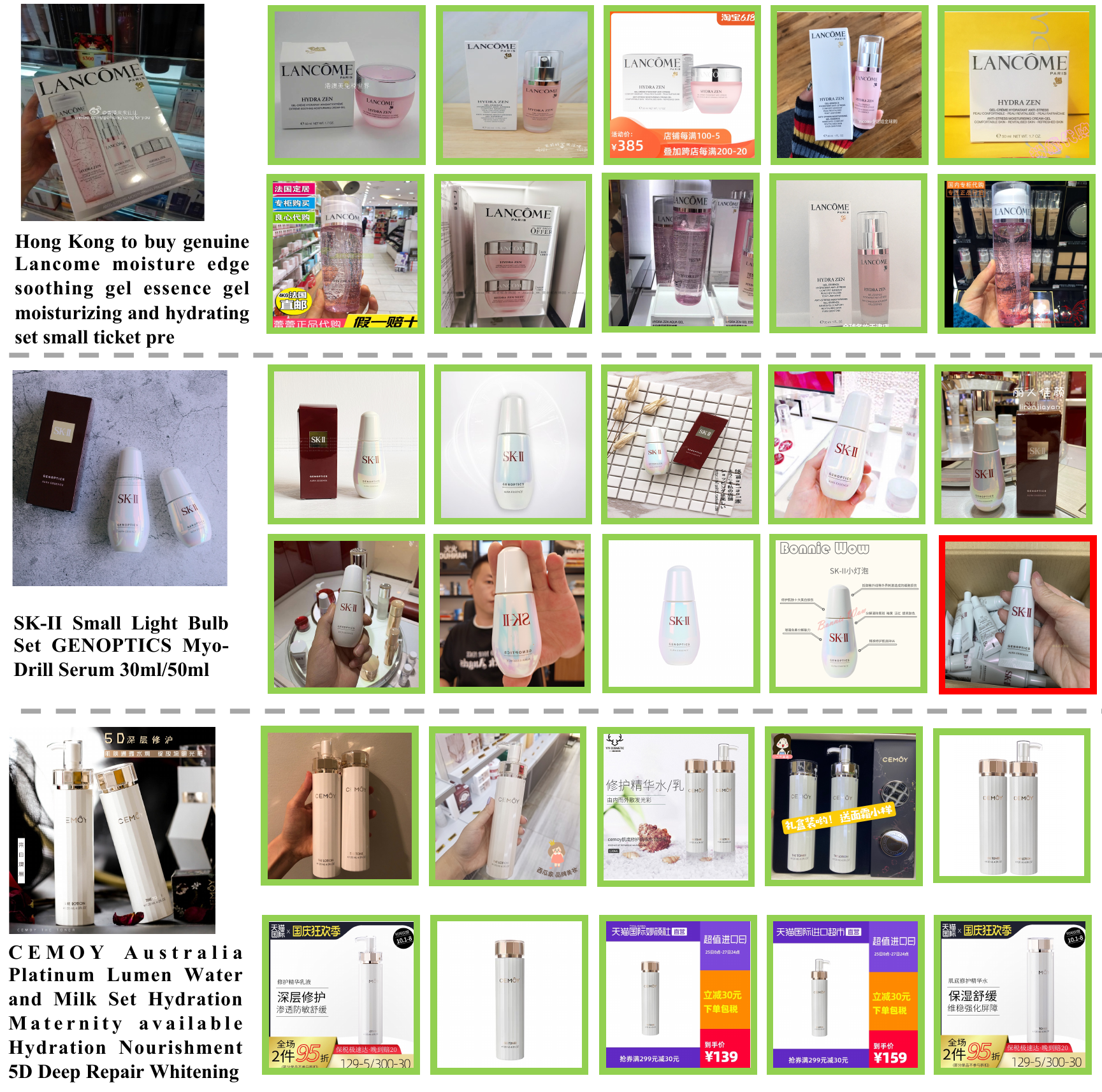}
    \caption{Visualizations of the EGE-CMP retrieval results.
    On the left are multi-product query images. Green/red boxes indicate Correct/Incorrect retrieval images. }
    \label{fig_visualize}
\end{figure}

\begin{figure*}
    \centering
    \includegraphics[width=1\textwidth]{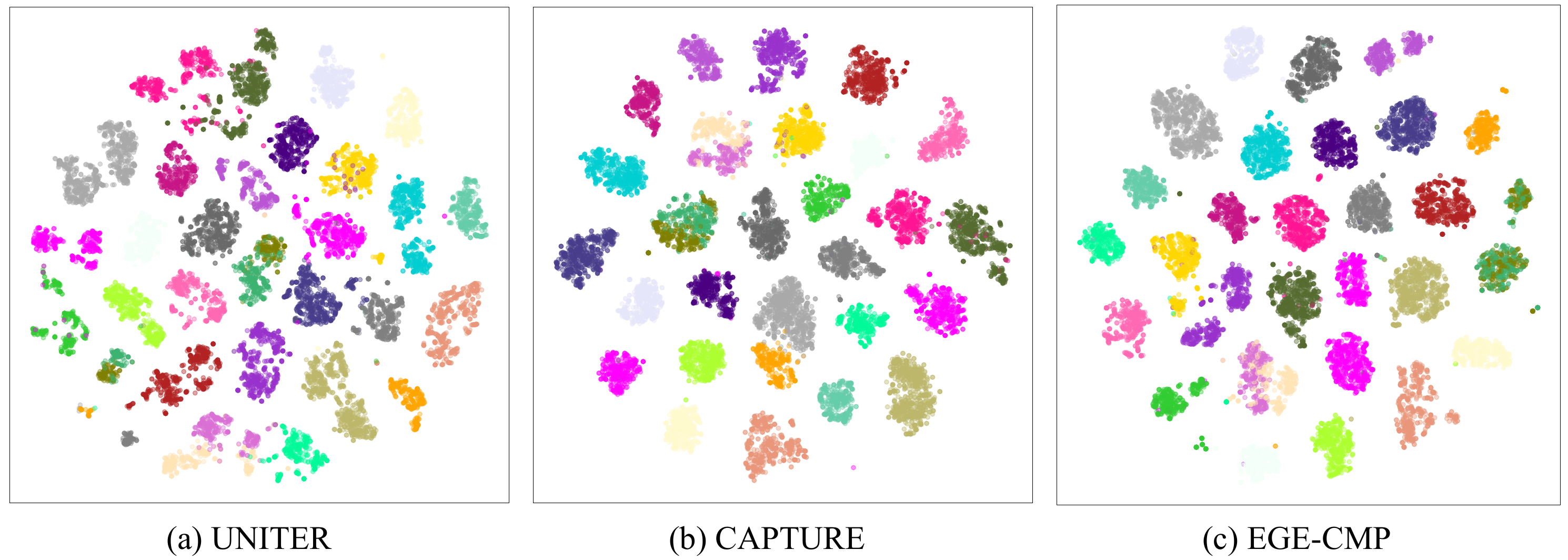}
    \caption{Visualize the embeddings generated by UNITER, CAPTURE and EGE-CMP via t-SNE. 
    Points belonging to the same category are of the same color. Best viewed in color.}
    \label{fig_embedding}
\end{figure*}

\subsection{Impact of Layer Configuration}
\label{sec_layer_ablation}

We study how the arrangement of transformer layers may affect the performance of our model in Table~\ref{tab:ablation_layer}.
In the Config column, the triplet represents the number of Text/Visual Transformer, Cross-Transformer, and Co-Transformer layers, respectively.
To provide a fair comparison, we first eliminate layers of a certain kind while maintaining the depth of the resultant network at the same level as EGE- CMP's, \textit{i.e.}, 12 levels.
The terms 'w/o-Cross', 'w/o-Co', and 'w/o-T/V/E' refer to the model obtained by deleting the EGE-CMP Cross-Transformer, Co-Transformer, and Text/Visual/Entity Transformer layers.
As can be observed, these three models perform worse than EGE-CMP, demonstrating the efficacy of its hybrid-stream design.
Additionally, we discover that a single model transformer (SMT) is critical for our model, demonstrating that a fine-tuned transformer is more beneficial for the overall model's performance.
Additionally, we investigate the combination of three-layer types in varying proportions in the second group of Table \ref{tab:ablation_layer} (EGE-CMP-A,B,C).
It turns out that the (4,4,4) combination gives the best performance.
We examine the performance of a smaller model (EGE-CMP-S) and a bigger model in greater detail (EGE-CMP-L).
As can be observed, EGE-CMP with the (4,4,4) configuration offers a more favorable trade-off between accuracy and parameters.

\subsection{Zero-Shot Instance-Level Retrieval}
We propose that a retrieval-based strategy is more applicable to real-world settings in which the category set is constantly updated and collecting vast volumes of clean labels is too expensive.
In contrast to detection, our retrieval approach does not rely on a fixed-size collection of predefined classes or on fine-grained box annotations.
To demonstrate this point, we perform zero-shot retrieval studies and provide the results in Table~\ref{tab:zero_shot}.
We manually remove 5/10/20 brands from the \textit{train} set and train EGE-CMP on the remaining samples to avoid discarding the deleted categories during training.
Then we compare EGE-CMP to the classes of these previously unknown brands.
Additionally, we compare our model to the LXMERT two-stream model and the UNITER single-stream model.
As can be observed, EGE-CMP outperforms LXMERT and UNITER on all three criteria, demonstrating its generalizability well.
Additionally, we display the embeddings created by UNITER, CAPTURE, and EGE-CMP in Figure~\ref{fig_visualize} using t-SNE~\cite{van2008visualizing}.
It turns out that the EGE-CMP characteristics are more discriminative, which is advantageous for the retrieval job.

\subsection{Comparisons on Single-Product Retrieval}
\label{sec_single_product_ablation}

It's worth noting that EGE-CMP may be used for both single-~\footnote{The single-product retrieval is a special case in the identical product search. There is only one instance in all product samples.} and multi-product retrieval.
Indeed, it excels at these two tasks and outperforms other baselines when it comes to single-product retrieval.
Specifically, we select a single-product sample from the \textit{gallery} set as a query and execute single-product retrieval from the \textit{gallery}set's remaining samples.
The performance of four models is compared in Table~\ref{tab:ablation_detector}, namely UNITER-single, LXMERT-single, CAPTURE-single, and EGE-CMP-single.
As can be shown, single-product retrieval performs far better than multi-product retrieval, as the difficulty is significantly decreased when there is only one instance/entity in the image/text.
Additionally, we see that the performance of 'EGE-CMP-single' is superior to that of two other baselines, demonstrating the superiority of EGE-CMP.


\section{Conclusion}\label{sec:conclu}
In this work, we describe the first attempt at extending the scope of canonical intra-/cross-modal retrieval to a more generalized scenario, namely, weakly-supervised multi-modal instance-level product retrieval, which has broad demands potential in E-commerce business.
To facilitate multi-modal pre-training, we present the Product1M, which is one of the biggest available visual-language retrieval datasets, as well as the only one designed exclusively for instance-level retrieval in real-world contexts. 
Additionally, we present an entity-graph enhanced hybrid-stream transformer, dubbed EGE-CMP, which excels at capturing potential synergy and understanding common relevant data from disparate modalities.
we incorporate genuine semantic information into the entity knowledge, focusing cross-modal contrastive learning on the related regions or words between multi-modal characteristics.
Extensive experimental evidence verifies that our EGE-CMP outperforms several state-of-art multi-modal pretraining models on a wide variety of metrics. 
We believe that the proposed EGE-CMP model, Product1M dataset, and established baselines will stimulate additional research, making it a more trustworthy and flexible retrieval search basement

\section{Acknowledgement}
This work was supported in part by Shenzhen Fundamental Research Program (Project No.RCYX20200714114642083,  No. JCYJ20190807154211365),  Guangdong Outstanding Youth Fund (Grant No. 2021B1515020061), National Key R$\&$D Program of China under Grant (No. 2018AAA0100300),  Guangdong Province Basic and Applied Basic Research (Regional Joint Fund-Key) Grant (No. 2019B1515120039), National Natural Science Foundation of China under Grant (No. U19A2073) and (No. 61976233).


\bibliographystyle{IEEEtran}
\bibliography{References}

\end{document}